# Dynamic Behavior in Piezoresponse Force Microscopy

Stephen Jesse, Arthur P. Baddorf, and Sergei V. Kalinin[*]

*Condensed Matter Sciences Division, Oak Ridge National Laboratory, Oak Ridge, TN 37831*

## Abstract

Frequency dependent dynamic behavior in Piezoresponse Force Microscopy (PFM) implemented on a beam-deflection atomic force microscope (AFM) is analyzed using a combination of modeling and experimental measurements. The PFM signal comprises contributions from local electrostatic forces acting on the tip, distributed forces acting on the cantilever, and three components of the electromechanical response vector. These interactions result in the bending and torsion of the cantilever, detected as vertical and lateral PFM signals. The relative magnitudes of these contributions depend on geometric parameters of the system, the stiffness and frictional forces of tip-surface junction, and operation frequencies. The dynamic signal formation mechanism in PFM is analyzed and conditions for optimal PFM imaging are formulated. The experimental approach for probing cantilever dynamics using frequency-bias spectroscopy and deconvolution of electromechanical and electrostatic contrast is implemented.

[*] Corresponding author, sergei2@ornl.gov

# I. Introduction

In the last decade, piezoresponse force microscopy was established as a powerful tool for probing local electromechanical activity on the nanometer scale.[1,2,3,4] Developed originally for imaging domain structures in ferroelectric materials, PFM was later extended to local hysteresis loop spectroscopy[5,6] and ferroelectric domain patterning for applications such as high density data storage[7,8] and ferroelectric lithography.[9,10,11] It was shown recently that vector PFM can be used to determine local molecular or crystallographic orientation in piezoelectric materials, provided that all three components of electromechanical response vector are determined quantitatively.[12] Broad applicability of PFM to materials such as ferroelectric perovskites, piezoelectric III-V nitrides,[13] and, recently, biological systems such as calcified and connective tissues,[14,15,16] has necessitated fundamental theoretical studies of image formation mechanism in PFM to provide the guidelines for quantitative data acquisition and interpretation. It was recognized that electrostatic tip-surface forces and buckling oscillations of the cantilever can provide significant and in some cases even dominating contributions to the PFM signal.[17,18,19] Imaging ferroelectric materials in the vicinity of a phase transition at small probing biases or imaging of biological systems with weak electromechanical coupling require optimal imaging conditions to be established, and a number of approaches based on using contact resonances in PFM have been suggested.[20,21] Finally, it is recognized that the use of the cantilever coupled with a beam-deflection detection system typical for most commercial AFMs does not allow longitudinal and normal force components to be unambiguously distinguished,[12,22] and it has been suggested that operation at specific frequencies would allow these components to be differentiated.[22] In our previous publications, we presented in-depth analysis of the static (low frequency) PFM imaging mechanism and demonstrated approaches for data interpretation and visualization.[12,23,24] In particular, we have shown that under the condition of good tip-surface contact (no potential drop in the tip-surface gap), materials properties measured by PFM are independent of the geometric characteristics of the tip, thus distinguishing this technique from mechanical SPM probes such as Atomic Force Acoustic Microscopy[25] or Ultrasonic Force Microscopy[26,27] that require that the tip shape be calibrated for quantitative measurements. This suggests that PFM is relatively insensitive to surface topography and provides quantitative information on material properties without the stringent requirement for tip shape calibration. However, to fully utilize PFM as a quantitative tool for local materials characterization, an understanding of the dynamic behavior, including the frequency dependent contrast in vertical and lateral PFM, is required.

Here, we analyze the dynamic behavior of cantilevers in PFM using both experimental bias-frequency-response diagrams and elastic beam theory. We analyze the difference between signal transduction in the vertical PFM (VPFM) and lateral PFM (LPFM) cases, discuss the contribution of longitudinal response to the VPFM signal and the contribution of cantilever buckling oscillations. The guidelines for quantitative PFM imaging and optimal frequency regimes are formulated.

## II. Principles of PFM

Piezoresponse Force Microscopy is based on the detection of bias-induced surface deformation.[3,4] The conductive tip is brought into contact with the surface and a periodic bias, $V_{tip} = V_{dc} + V_{ac} \cos(\omega t)$, is applied. The piezoelectric response of the surface is detected as a periodic deflection of the cantilever due to the first harmonic component, $d_{1\omega}$, of the bias-induced surface displacement, $d = d_0 + d_{1\omega} \cos(\omega t + \varphi)$. On a beam-deflection system, an experimentally measured PFM signal is related to the voltage generated at the photodiode. In most cases, the vertical sensitivity is calibrated so that piezoresponse amplitude, $A$, is measured in the units of height. For a purely vertical surface displacement in the low frequency regime, $A = d_{1\omega}$. However, for realistic cases, the relationship between the surface displacement vector and the measured vertical PFM signal is more complex, as will be discussed below.

Provided that response is a linear function of driving amplitude, the piezoresponse amplitude, $PR = A/V_{ac}$, defines the local electromechanical activity of the surface. The phase of the electromechanical response of the surface, $\varphi$, yields information on the polarization direction below the tip. For a purely electromechanical response, the piezoresponse amplitude is equal for $c^+$ and $c^-$ domains, while the phase changes by 180° between the domains.

One of the major complications in PFM is that both long range electrostatic forces and the electromechanical response of the surface contribute to the PFM signal so that the experimentally measured piezoresponse amplitude is $A = A_{el} + A_{piezo} + A_{nl}$, where $A_{el}$ is an electrostatic contribution due to tip-surface forces, $A_{piezo}$ is an electromechanical contribution due to piezoelectric surface deformation and $A_{nl}$ is a non-local contribution due to capacitive cantilever-surface interactions.[17,28] For most electromechanically active materials with dominant piezoelectric coupling, (excluding e.g. PZT compositions close to a morphotropic phase boundary or electrostrictive polymers), the electromechanical response is linear in bias, $A_{piezo} = d_{eff}(V_{dc})V_{ac}$, i.e., the electrostrictive contribution is small. For linear piezoelectric materials, $d_{eff}(V_{dc}) = const$ is a bias-independent piezo-coefficient. However, for ferroelectric materials, the bias dependence of $d_{eff}(V_{dc})$ is determined by the local polarization switching processes and the local electromechanical hysteresis loop, which can be measured by PFM spectroscopy. Local electrostatic contributions vary linearly with the tip-surface potential difference, $A_{el} = G_{loc}V_{ac}(V_{dc} - V_{surf})$, where $V_{surf}$ is position dependent local surface potential and $G_{loc} = dC_{loc}/dz$ is tip-surface capacitance gradient. Due to the fact that the cantilever size is much larger than a typical feature size, the non-local contribution can be well-approximated as being position independent, $A_{el} = G_{nl}V_{ac}(V_{dc} - V_{av})$, where $V_{av}$ is average surface potential and $G_{nl} = dC_{nl}/dz$ is the cantilever-surface capacitance gradient. Quantitative PFM imaging requires $A_{piezo}$ to be maximized in order to achieve a predominantly electromechanical signal. Contribution of the non-local electrostatic term to the PFM contrast was analyzed by several authors and it was suggested that a phase shift of 180° between the domains is an indication of purely electromechanical imaging.[3,4] However, to date such analyses have ignored the dynamic properties of the cantilever. Below we demonstrate that this conjecture can be validated through the analysis of dynamic cantilever properties.

Electrostatic contributions to the PFM signal can be minimized by a suitable choice of dc tip potential, $V_{dc}$. However, for ferroelectric materials, the local surface potential is related to the domain structure.[29] For other materials, the presence of surface contaminates and remnant charges can often result in significant potential variations across the surface. It has been shown that surface potential and PFM data can be acquired simultaneously using lift mode,[29] thus potentially paving the way for techniques in which the dc potential correction is position dependent. However, the difference in spatial resolution between PFM (~5 nm) and potential-sensitive techniques such as Scanning Surface Potential Microscopy (~50-100 nm) and, more importantly, the PFM spectroscopy requirement for $V_{dc}$ to be varied, render this approach non-universal. Therefore, minimization of the electrostatic contribution requires an optimal probe geometry and operational frequency, as will be analyzed below.

One of the difficulties in comparing the relative magnitudes of electromechanical and electrostatic responses is the difference in the contrast transfer mechanism. In the electromechanical case, the surface displacement due to the inverse piezoelectric effect is determined as a function of the applied voltage. For low frequencies, tip deflection is equal to surface displacement since the contact stiffness of the tip–surface junction is usually much larger than the cantilever spring constant. In the electrostatic case, the force containing both local and non-local components is a primary effect, and force-induced indentation is determined by the tip radius and the tip-surface junction stiffness.

An additional level of complexity is added in the PFM imaging of materials that are not transversally isotropic. In this case, the electromechanical response of the surface is a vector with non-zero in-plane components. An approach to access the electromechanical properties of such materials has been introduced by the development of lateral PFM, in which the torsional component of bias induced tip vibration is detected.[30,31] This approach was further extended to 3D PFM imaging, in which both in-plane components of the response vector are determined with the out-of-plane component.[30,32] Note that measurement of the electromechanical response vector, as opposed to the qualitative data, is impossible unless the vertical and lateral signals are properly calibrated, i.e. measured in the same units. Provided that relative sensitivities of the vertical and lateral PFM signals are known (e.g. Ref. [33]), the complete electromechanical response vector can be measured, an approach referred to as Vector PFM.[12] Transduction of surface vibrations to the tip is significantly different for vertical and in-plane components. For the latter, the onset of sliding between the tip and the surface can affect the transduction. This mechanism is absent in the vertical direction.

In addition to vertical surface oscillations, in-plane surface displacement along the cantilever axis (longitudinal displacement) contributes to the cantilever bending and thus to PFM contrast. This mechanism is illustrated in Fig. 2d, in which the vertical displacement of the cantilever is measured through the deflection angle of the cantilever, $\theta_d \sim w_1/L$, where $L$ is cantilever length. Surface displacement along the cantilever axis will also result in the change of deflection angle. If the surface displacement and the tip displacement are equal (no slip condition), the deflection angle would be $\theta_d = w_2/H$, where $H$ is the tip height. Given that typically $H \ll L$, this implies that the "vertical" PFM signal is more sensitive to the longitudinal surface displacement than to the vertical surface displacement. This is also the case for amplification between flexural and torsional modes, as noted by Peter et al.[33]. However, in the longitudinal and torsional cases, the cantilever can slip along the surface, resulting in a different signal transduction mechanism, which reduces the impact of the longitudinal displacement on the deflection angle.

These considerations illustrate that dynamic image formation in PFM is very complex due to a variety of contributions to the signal (distributed and localized electrostatic forces, and surface displacement) and the signal transduction mechanism (elastic coupling, friction) between the tip and surface. Below, we study theoretically the dynamic behavior of the cantilever in vertical and lateral PFM, and analyze the implications of this behavior on PFM imaging and quantification.

## III. Theory

A cantilever in combination with an optical beam deflection detector is the key part of the SPM force detection mechanism. The motion of the cantilever induced by surface oscillations has been studied extensively in the context of Atomic Force Acoustic Microscopy (AFAM)[34,35,36,37] and Ultrasonic Force Microscopy (UFM).[38] However, electrostatic modulation in PFM gives rise to additional local and non-local force contributions that can couple to the displacement induced oscillations. The resonant frequencies corresponding to different vibration modes (e.g. torsional and vertical oscillations) are generally different. In most SPM techniques, ranging from conventional intermittent contact mode topographic imaging to AFAM and UFM, determining optimal imaging condition requires selecting the proper modulation frequency. Hence, analysis of cantilever dynamics is required in order to determine whether this approach can be used to selectively amplify a particular component of the PFM signal, and thereby provide information on vertical and lateral electromechanical responses and electrostatic properties of the surface. As discussed above, analysis of the dynamic image formation mechanisms in vector PFM should necessarily take into account the following contributions as illustrated in Figs. 1 and 2:

1. The local vertical surface displacement translated to the tip

2. The longitudinal, in-plane surface displacement along the cantilever axis

3. The lateral surface displacement, in-plane and perpendicular to the cantilever axis

4. The local electrostatic force acting on the tip

5. The distributed electrostatic force acting on the cantilever

Below, we analyze the dynamic behavior of the probe in these five cases. The cantilever is modeled as a uniform beam parallel to the surface. The use of more rigorous models, including cantilever tilt, can improve this model. However, in the general case, the surface slope is unknown, and the solution, while being amenable to analytical or numerical methods, is not universal and is significantly more cumbersome, precluding insight into the imaging mechanism. For clarity, we also neglect damping, which can be easily incorporated, but the solutions become more complex.

Bias-dependent contact mechanics of the tip-surface junction is considered in Section III.1. The general solution for cantilever dynamics in PFM is derived in Section III.2. In Section III.3, considered is the important case of cantilever vibration under purely electrostatic forces, providing an approach for probe calibration. Sections III.4 and III.5. discuss the limiting cases of zero and infinite lateral stiffness constant, corresponding to frictionless contact and cantilevered piezoindentation. Finally, torsional response of the cantilever is briefly analyzed in Section III.6.

### III.1. Contact mechanics and boundary conditions

The contrast formation mechanism in PFM is determined by the interplay of contact mechanics of the tip-surface junction and cantilever dynamics. The mechanical equivalent circuit can be represented by two springs, connected in series, having spring constants $k_1$ and $k_2$, as shown in Fig. 3. Local electromechanical contributions to the PFM signal arise due to the bias induced surface displacement, represented as $d_1$ and $d_2$. Note that for the cantilever based force sensor, vertical and lateral contact mechanics are coupled, and even for a purely vertical PFM signal, the tip can shift along the surface.

In the Hertzian approximation[39] for a spherical tip, the vertical surface to tip spring constant, $k_1 = 1.82(E^*)^{2/3} R^{1/3} P^{1/3}$, where $P$ is indentation force, $R$ is tip radius of curvature and $E^*$ is the indentation modulus. The indentation force is $P = kA_0$, where $k$ is the spring constant of the cantilever and $A_0$ is the static set-point cantilever deflection. The indentation mechanics for piezoelectric materials is more complex, and an exact solution is available only for transversally isotropic piezoelectric materials.[23,24] In this case, the spring constant of the tip-surface junction is bias dependent, $k_1 = R^{1/2}\left(2w_0^{1/2}C_1^* - w_0^{-1/2}V_{tip}C_3^*\right)/\pi$, where $w_0$ is an indentation depth determined by the stiffness relation $P = \left(4w_0^{3/2}R^{1/2}C_1^* + 6w_0^{1/2}R^{1/2}V_{tip}C_3^*\right)/3\pi$. For a typical ferroelectric, such as BaTiO$_3$, in the $c^+$ domain state, with an indentation elastic modulus of $C_1^* = 403\,\text{GPa}$, an indentation piezoelectric modulus of $C_3^* = 15.4\,\text{N/Vm}$, a tip radius of $R = 50$ nm, an applied force of $P = 100$ nN, the indentation depth is $w_0 = 3.01\text{A}$, and the effective tip-surface spring constant is $k_1 = (993 - 63.3 V_{tip})\,\text{N/m}$. The bias dependence of the tip-surface spring constant is relatively weak and becomes even smaller for a flattened tip that is typically used in AFAM and UFM.[27] The contact spring constant, $k_1 \sim 1000\,\text{N/m}$, is significantly higher than the typical cantilever spring constant $k \sim 1-50\,\text{N/m}$. Thus, in the low frequency regime, the vertical tip displacement can be found as $\delta A = k_1 \delta w/(k + k_1)$, where $\delta w = d_{eff} V_{ac}$ is the bias induced surface displacement, $k$ is the cantilever spring constant, and $k_1$ is the vertical spring constant of the tip-surface junction. Hence, the tip deflection is almost equal to the surface displacement, $\delta A \approx \delta w$, which is the usual assumption in PFM. However, similarly to AFAM and UFM, this approximation is not valid in the high-frequency regime above the first resonant frequency of the cantilever, where inertial stiffening effects become important.

In the linear model, the lateral spring constant of the tip-surface junction is $k_2 = 8aG^*$, where $a$ is the contact radius and $G^*$ is the shear modulus. For isotropic materials, the ratio between the two is determined only by the Poisson module and is $k_1/k_2 = 2(1-\nu)/(1-2\nu) \approx 0.85$ for most materials. The voltage dependence of the lateral spring constant for piezoelectric materials has not, to our knowledge, been addressed yet.

The local electrostatic contribution to the deflection signal arises due to the force-induced indentation of the surface, where the force, $F_{loc}$, is the first harmonic of the capacitive force between the conical and spherical part of the tip and the surface. In the low frequency regime, $A_{el} = F_{loc}/k_1$. For typical PFM imaging conditions, the first harmonic of the local electrostatic force contribution was estimated as $F_{loc} = 1.4 \cdot 10^{-8}$ N/V for a tip-surface separation of z = 0.1 nm and tip radius R = 50 nm.[28] Thus, the electrostatic contribution, $A_{el} = 14 V_{tip}$ (pm), can be comparable to the electromechanical contribution, depending on the imaging conditions and tip geometry. The response diagrams relating dominant contrast mechanisms to experimental parameters in the low frequency regime are reported elsewhere.[28] Similarly to electromechanical contributions, dynamic effects in the high-frequency regime will strongly affect the magnitude of local and distributed electrostatic responses.

The non-local contribution to PFM, $A_{nl}$, arises due to the buckling oscillations of the cantilever induced by capacitive cantilever-surface interactions as illustrated in Figure 1b and will be analyzed in detail in Section V. The relative magnitudes of these five contributions are frequency dependent as analyzed below.

### III.2. Cantilever dynamics in PFM

The dynamic behavior of the cantilever in the general case can be described by the beam equation

$$\frac{d^4 u}{dx^4} + \frac{\rho S_c}{EI}\frac{d^2 u}{dt^2} = \frac{q(x,t)}{EI}, \quad (1)$$

where $E$ is the Young's modulus of cantilever material, $I$ is the moment of inertia of the cross-section, $\rho$ is density, $S_c$ is cross-section area, and $q(x,t)$ is the distributed force acting on the cantilever. For a rectangular cantilever $S_c = wh$ and $I = wh^3/12$, where $w$ is the cantilever width and $h$ is thickness. The cantilever spring constant, $k$, is related to the geometric parameters of the cantilever by $k = 3EI/L^3 = Ewh^3/4L^3$. In beam-deflection SPM, the deflection angle of the cantilever, $\theta$, is measured by the deflection of the laser beam at $x = L$, and is related to the local slope as $\theta = \arctan(u_0'(L)) \approx u_0'(L)$. For a purely vertical displacement, the relationship between cantilever deflection angle and measured height is $A = 2\theta L/3$.[40] Thus, in cases when the deflection angle is determined by either longitudinal or electrostatic contributions, the effective vertical displacement measured by AFM electronics will also be related to the deflection angle as $A = 2\theta L/3$.

Eq. (1) is solved in the frequency domain by introducing $u(x,t) = u_0(x)e^{i\omega t}$, $q(x,t) = q_0 e^{i\omega t}$, where $u_0$ is the displacement amplitude, $q_0$ is a uniform load per unit length, $t$ is time, and $\omega$ is modulation frequency. After substitution, Eq. (1) is:

$$\frac{d^4 u_0}{dx^4} = \kappa^4 u_0 + \frac{q_0}{EI}, \quad (2)$$

where $\kappa^4 = \omega^2 \rho S_c/EI$. On the clamped end of the cantilever, the displacement and deflection angle are zero, yielding the boundary conditions

$$u_0(0) = 0 \text{ and } u_0'(0) = 0, \quad (3a,b)$$

On the supported end, in the limit of linear elastic contact the boundary conditions for moment and shear force are

$$EIu_0''(L) = k_2 H\left(\tilde{d}_2 - u_0'(L)H\right) \text{ and } EIu_0'''(L) = -f_0 + k_1\left(u_0(L) - d_1\right) \quad (4a,b)$$

where is $d_1 = d_{vert} V_{ac}$ is the first harmonic component of bias-induced vertical surface displacement due to the piezoelectric effect, $d_2 = d_{lat} V_{ac}$ is the first harmonic component of the longitudinal surface displacement, $f_0$ is the first harmonic of the local force, $f(x,t) = f_0 e^{i\omega t}$, acting on the tip, and $k_1$ and $k_2$ are the vertical and longitudinal spring constants of the tip-surface junction (Fig. 3). For non-piezoelectric materials, $d_1 = d_2 = 0$, while for zero electrostatic force, $f_0 = 0$, providing purely electromechanical and purely electrostatic limiting cases for Eq. (2).

Because Eq. (2) is linear, it can be solved in the usual fashion. Using $EI = kL^3/3$, the dynamic behavior of the cantilever is given by

$$\theta_{tot} = \frac{A_v(\beta)d_1 + A_l(\beta)d_2 + A_e(\beta)f_0 + A_q(\beta)q_0}{N(\beta)} \quad (5)$$

where

$$A_v(\beta) = 3\beta^4 k_1 kL \sin\beta \sinh\beta \quad (6)$$

$$A_l(\beta) = 3\beta^2 Hk_2 \left(3k_1 + \cosh\beta\left(-3k_1\cos\beta + \beta^3 k\sin\beta\right) + \beta^3 k\cos\beta\sinh\beta\right) \quad (7)$$

$$A_e(\beta) = 3\beta^4 kL \sin\beta \sinh\beta \quad (8)$$

$$A_q(\beta) = 3L^2\left(3k_1(\cos\beta - \cosh\beta) - k\beta^3\sin\beta + (k\beta^3 + 3k_1\sin\beta)\sinh\beta\right) \quad (9)$$

$$N(\beta) = \beta^2(9H^2 k_1 k_2 + \beta^4 k^2 L^2 + \cosh\beta((-9H^2 k_1 k_2 + \beta^4 k^2 L^2)\cos\beta + \\ + 3\beta k(k_1 L^2 + H^2 k_2 \beta^2)\sin\beta) + 3\beta k(-k_1 L^2 + H^2 k_2 \beta^2)\cos\beta\sinh\beta) \quad (10)$$

and the dimensionless frequency is $\beta = \kappa L$.

The ratios $A_v(\beta)/N(\beta)$, $A_l(\beta)/N(\beta)$, $A_e(\beta)/N(\beta)$, $A_q(\beta)/N(\beta)$ describe the frequency dependence of the PFM signal due to vertical and longitudinal components of surface displacement, the local electrostatic force acting on the tip, and the distributed electrostatic force acting along the cantilever, respectively. Note that the vertical electromechanical contribution and local force contribution have similar frequency dependences (compare Eqs. (6) and (8)).

The resonance structure in Eq. (5) is determined only by the properties of the cantilever and the spring constant of the tip-surface junction. Thus, the resonance frequencies corresponding to the roots of Eq. (10) are independent of the relative contributions of electrostatic and electromechanical interactions. Therefore, tracking the resonant frequency of electrically excited cantilever as a function of tip position provides information on local elastic properties, which is similar to frequency detection in AFAM. The dependence of resonant frequencies on the vertical and lateral spring constants of the tip-surface junction is analyzed in detail by Rabe.[41] Moreover, since the denominator of Eq.(5) does not depend on the relative magnitudes of vertical, longitudinal, and electrostatic responses, these contributions cannot be separated by a proper choice of driving frequency. Therefore, unambiguous measurement of all three components of the electromechanical response vector requires alternative solutions, e.g. based on either 3D SPM[42] or sample rotation.[30] At the same time, electromechanical, local, and non-local contributions to the PFM are additive, making it possible to distinguish the relative contributions of these signals to the observed contrast.

In the low frequency limit, $\beta \to 0$, these contributions become:

$$\frac{A_v(\beta)}{N(\beta)} = \frac{6k_1 kL}{4k(k+k_1)L^2 + 3H^2 k_2 (4k+k_1)} \approx \frac{6kL}{4kL^2 + 3H^2 k_2} \tag{11}$$

$$\frac{A_l(\beta)}{N(\beta)} = \frac{3Hk_2(4k+k_1)}{4k(k+k_1)L^2 + 3H^2 k_2 (4k+k_1)} \approx \frac{3Hk_2}{4kL^2 + 3H^2 k_2} \tag{12}$$

$$\frac{A_e(\beta)}{N(\beta)} = \frac{6kL}{4k(k+k_1)L^2 + 3H^2 k_2 (4k+k_1)} \approx \frac{6kL}{k_1(4kL^2 + 3H^2 k_2)} \tag{13}$$

$$\frac{A_q(\beta)}{N(\beta)} = \frac{(8k-k_1)L^2}{16k(k+k_1)L^2 + 12H^2 k_2(4k+k_1)} \approx \frac{L^2}{4(4kL^2 + 3H^2 k_2)} \tag{14}$$

Eqs. (11-14) thus describe static deflection at frequencies well below the first resonance. Note that the relative magnitudes of these contributions are, as expected, sensitive to the length of the cantilever and for stiff cantilevers, $kL^2 >> H^2 k_2$, the response is determined primarily by the vertical displacement of the tip. For a soft cantilever, the in-plane displacement of the tip apex is the dominant oscillation mode. From the equations (11-14), the scaling of the electromechanical and electrostatic responses with the geometric parameters of the cantilever ($L$, $H$) and corresponding spring constants ($k$, $k_1$, $k_2$) can be determined in a straightforward manner.

From Eqs. (5-10), the frequency dependence of the non-local electrostatic, local electrostatic, and piezoelectric contributions to be estimated as

$$\frac{A_v(\beta)}{N(\beta)} \sim \frac{k_1}{\kappa^2} \sim \frac{k_1}{\omega} \tag{15}$$

$$\frac{A_l(\beta)}{N(\beta)} \sim \frac{1}{\kappa} \sim \frac{1}{\omega^{1/2}} \tag{16}$$

$$\frac{A_e(\beta)}{N(\beta)} \sim \frac{1}{\kappa^2} \sim \frac{1}{\omega} \tag{17}$$

$$\frac{A_q(\beta)}{N(\beta)} \sim \frac{1}{\kappa^3} \sim \frac{1}{\omega^{3/2}} \tag{18}$$

From Eqs. (15-18), all four contributions decrease with frequency due to the dynamic stiffening effects. Even in the absence of damping, the non-local contribution scales as a higher power of frequency, suggesting that non-local cantilever effects will be minimized at high frequencies. At the same time, the local electrostatic and electromechanical contributions scale in a similar manner as the ratio, $A_{piezo}/A_{el} = d_1 k_1/f_0$, which depends only on the spring constant of the tip-surface junction. This suggests that these contributions cannot be distinguished by a choice of the operating frequency. Instead, either the use of a cantilever with a high spring constant ($k_1 \to \infty$) or imaging at the nulling bias or using shielded probes[43] (where $f_0 \to 0$) is required. In the linear elastic approximation, the dynamic stiffening effect is less pronounced for the longitudinal signal; however, in this case, the onset of sliding friction can minimize the longitudinal effect. This behavior is analyzed in more detail in Section IV.2.

The general frequency dynamics of PFM, as described by Eq. (5), are extremely complex and sensitively depend on vertical and longitudinal spring constants of the tip-surface junction and geometric parameters of the cantilever. However, a number of important limiting cases can be selected, as summarized below.

### III.3. Cantilever dynamics under electrostatic forces

One important limiting case is that of the pure buckling oscillations of the cantilever due to a distributed electrostatic force. In this case, $k_1 \to \infty$ and $k_2 \to \infty$ in Eq. (5), and the boundary conditions on the end supported by the probe tip in contact with the surface can be written as:

$$u_0(L) = 0 \text{ and } u_0''(L) = 0, \tag{19a,b}$$

corresponding to zero displacement and no applied rotational moment. The frequency response of such a supported cantilever is

$$\theta_{nlc} = \frac{3q_0}{k\beta^3} \frac{\cos\beta - \cosh\beta + \sin\beta \sinh\beta}{\cosh\beta \sin\beta - \cos\beta \sinh\beta} \tag{20}$$

The resonant frequencies are found as the roots of denominator, $\cosh\beta \sin\beta = \cos\beta \sinh\beta$. The lowest order resonances in Eq. (20) are to $\beta_n$ = 3.927, 7.067, 10.21. The corresponding eigen frequencies are $\omega_n^2 = EI\beta_n^4/\rho S_c L^4 = Eh^2\beta_n^4/12\rho L^4$. In terms of the cantilever spring constant, $k_c = Ewh^3/4L^3$, the resonant frequencies can be found to be $\omega_n^2 = k\beta_n^4/3m$, where $m = hLw\rho$ is cantilever mass. In the low frequency regime, $\theta_{nlc} = q_0/16k$, resulting in a measured vertical response of $A = q_0L/24k$.

In comparison, for a free cantilever (tip not touching the surface) oscillating under the simultaneous action of distributed and localized forces, the boundary conditions on the free end are

$$u_0''(L) = 0 \text{ and } EIu_0'''(L) = -f_0, \tag{21 a,b}$$

where $f_0$ is the first harmonic component of the electrostatic force acting on the tip. The cantilever dynamic response in this case is

$$\theta_f = -\frac{-3Lq_0 \sin\beta + 3(Lq_0 + \beta f_0 \sin\beta)\sinh\beta}{k\beta^3 L(1 + \cosh\beta \cos\beta)} \tag{22}$$

At low frequencies, the static cantilever deflection is $\theta_f = (q_0/2k + 3f_0/2kL)$ and therefore, the measured height signal is $A = (q_0 L/3k + f_0/k)$. For a free cantilever, the response due to the non-local force is 8 times larger than that for the clamped cantilever.

The resonances in Eq. (22) occurs when $\cosh\beta\cos\beta + 1 = 0$. Several of the lowest order roots are $\beta_n$ = 1.875, 4.694, 7.855. Thus, the first cantilever buckling resonance in contact mode occurs at ~ 4.4 times the frequency of the resonance of the free cantilever for a distributed force.

The resonant frequencies for the electrostatically driven free cantilever are independent of the ratio between the distributed and localized forces. Currently, the approaches for the calibration of the cantilever spring constant are almost universally based on the determination of the resonant frequencies of the free cantilever[44,45,46,47,48] typically determined using mechanical excitation on the clamped end. From this analysis, we conclude that electrostatic excitation in the non-contact regime can be used to determine resonant frequencies as well, thus providing a convenient approach for cantilever calibration for PFM measurements.

### III.4. Frictionless contact

An important limiting case for probe dynamics in PFM corresponds to frictionless contact between the tip and the surface, $k_2 \rightarrow 0$. This can also be obtained by changing the boundary condition given in Eq. (4a) to $u_0''(L) = 0$. In this case, the dynamic response is given by

$$A_v(\beta) = 3\beta^4 k_1 k \sin\beta \sinh\beta \tag{23a}$$

$$A_l(\beta) = 0 \tag{23b}$$

$$A_e(\beta) = 3\beta^4 k \sin\beta \sinh\beta \tag{23c}$$

$$A_q(\beta) = 3L\big(3k_1(\cos\beta - \cosh\beta) - k\beta^3 \sin\beta + (k\beta^3 + 3k_1 \sin\beta)\sinh\beta\big) \tag{23d}$$

$$N(\beta) = \beta^3 kL\big(\beta^3 k + \cosh\beta(\beta^3 k \cos\beta + 3\beta k_1 \sin\beta) - 3k_1 \cos\beta \sinh\beta\big) \tag{23f}$$

If, additionally, the vertical spring constant of the tip-surface junction is large compared to that of the cantilever, $k_1 \to \infty$, Eq. (23) simplifies as

$$\theta_p = \frac{3Lq_0(\cos\beta - \cosh\beta) + (d_1 k\beta^4 + Lq_0)\sin\beta \sinh\beta}{\beta^3 kL(\cosh\beta \sin\beta - \cos\beta \sinh\beta)} \tag{24}$$

Eq. (24) describes the cantilever dynamics for the case of a soft cantilever on a piezoelectric surface under the effect of both a piezoelectric deformation and a distributed load.

In a more realistic case, when the cantilever can slide along the surface, the friction force will be non-zero. The rigorous description of this process for a periodic surface motion is very complex, and here we consider only the limiting case of linear sliding friction. The boundary conditions on the clamped end are given by Eq. (3a,b), whereas, on the supported end,

$$EIu_0''(L) = f_l H \text{ and } u_0'''(L) = 0 \tag{25a,b}$$

where the friction force is $f_l = \mu P$, $\mu$ is a friction coefficient and $P$ is the static vertical load. The solution for $q_0 = 0$ in this case is

$$\theta_f = \frac{3f_l H(\cosh\beta \sin\beta + \cos\beta \sinh\beta)}{\beta kL^2(1 + \cos\beta \cosh\beta)} \tag{26}$$

The resonances are similar to that of a free cantilever under the electrostatic forces given in Eq. (20). Note that in a realistic case the lateral surface oscillation amplitude in PFM is small (< 1 nm), and Eq. (5) is expected to be applicable in the low frequency regime. The dynamic behavior of the cantilever in the regime when the tip can slide along the surface is very complex and requires quantitative models for lateral contact mechanics of the tip surface junction.[49] Here, these effects will be addressed experimentally in Section V. Also note that the inclination of the surface with respect to cantilever will lead to coupling between normal and lateral contributions, e.g. as analyzed by Rabe for AFAM.[41,50]

### III.5. Piezoelectric nanoindentation

In the case when only the vertical displacement at the tip-surface junction is considered (stiff cantilever, $kL^2 >> H^2 k_2$) and the distributed force is ignored, the general solution for probe displacement is:

$$u(L) = \frac{3(f_0 + d_1 k_1)(\cos\beta \cosh\beta - 1)}{\beta^3 k + \cosh\beta(3k_1 \sin\beta - \beta^3 k \cos\beta) - 3k_1 \cos\beta \sinh\beta}, \quad (27)$$

and describes the signal detected by a nanoindentation experiment with interferometric detection. For PFM with beam-deflection detection, the angle of the tip is given by:

$$\theta_p = \frac{3\beta(f_0 + d_1 k_1)\sin\beta \sinh\beta}{L(\beta^3 k + \cosh\beta(\beta^3 k \cos\beta + 3k_1 \sin\beta) - 3k_1 \cos\beta \sinh\beta)} \quad (28)$$

Eq. (28) thus describes the frequency dynamics in the most common case of the PFM experiment performed with stiff cantilever on beam-deflection system. For low frequencies, Eq. (28) becomes $\theta_{p0} = 3(f_0 + d_1 k_1)/(2k_1 + 2k)L$. In this limit, corresponding to an infinitely stiff contact, $k_1 \to \infty$, the response becomes $\theta_{f0} = 3d_1/2L$, i.e. a purely electromechanical response. By using the relationship between height and deflection angle, $A = 2\theta L/3$, the measured signal is equal to surface displacement, $A = d_1$, thus recovering the usual assumption in vertical PFM. For a free cantilever, $k_1 \to 0$, the response is $\theta_{f0} = -3f_0/2kL$ and $A = -f_0/k$, i.e. a purely electrostatic response. In the contact regime for non-piezoelectric material, $\theta_{p0} = -3f_0/2k_1 L$ and the measured deflection is $A = -f_0/k_1$, i.e. measured displacement scales reciprocally with the tip-surface junction spring constant and is independent of the cantilever length, as expected. For non-zero frequencies in the limit $k_1 \to \infty$, Eq.(28) becomes

$$\theta_f = \frac{-d_1 \kappa \sin\beta \sinh\beta}{\cosh\beta \sin\beta - \cos\beta \sinh\beta}, \qquad (29)$$

whereas for $k_1 \to 0$, Eq. (28) becomes Eq. (5) for a free cantilever. Eq. (29) thus describes the contact dynamics of PFM in the purely electromechanical case.

Note that the resonant frequencies for the cantilever increase with the spring constant of the tip-surface junction, indicative of the stiffening of the corresponding mode. The individual branches are separated by frequency gaps, as has been shown by Rabe.[34] The crossover between models for free and supported cantilever shifts to high contact spring constants with the resonance number (dotted line in Fig. 4b), indicative of the dynamic stiffening of the cantilever. Thus, imaging at higher resonant frequencies enhances the sensitivity to elastic properties of materials with high elastic moduli.

### III.6. Lateral Contribution

In lateral PFM, torsional oscillations of the cantilever induced by the surface displacement are detected. The distributed and localized electrostatic forces are generally directed along the surface normal and are symmetric with respect to the cantilever and therefore can be expected to provide minimal contribution to the torsional behavior, i.e. these degrees of freedom are decoupled. The torsional oscillations have been extensively studied in the context of frictional acoustic force microscopy.[51,52] Here, we briefly repeat this analysis for PFM.

The equation for torsional oscillations is

$$c_T \frac{\partial^2 \Theta}{\partial x^2} = \rho J \frac{\partial^2 \Theta}{\partial t^2} \tag{30}$$

where $\Theta$ is the angle of torsion and $c_T$ is the torsional stiffness. For a rectangular beam, $c_T = wh^3 G/3$, where $G$ is the shear modulus of cantilever material. In the frequency domain, Eq. (30) becomes

$$\frac{\partial^2 \Theta_0}{\partial x^2} + \xi^2 \Theta_0 = 0 \tag{31}$$

where $\xi^2 = \rho J \omega^2 / c_T$ and $\Theta(x,t) = \Theta_0(x) e^{i\omega t}$. The boundary condition on the clamped end is $\Theta_0(0) = 0$. On the supported end, the boundary conditions are obtained in a similar way as for longitudinal oscillations. For linear elastic contact, the boundary condition on the supported end is

$$\Theta_0'(L) = -\tilde{k}_2 (H\Theta_0 - d_3) \tag{32}$$

where $d_3$ is the in-plane displacement and $\tilde{k}_2 = H k_2 / c_T$. The solution in this case becomes

$$\Theta_0(L) = \frac{d_3 \tilde{k}_2 \sin \xi L}{\xi \cos \xi L + H \tilde{k}_2 \sin \xi L} \tag{33}$$

For an infinitely stiff contact, $\tilde{k} \to \infty$, $\Theta_0(L) = \tilde{d}_3 / H$, while for a free cantilever, $k_2 \to 0$ and $\Theta_0(L) = 0$, as expected. Note that the resonant frequencies, as determined by the denominator of Eq. (33), depend on the stiffness constant, $k_2$ (Fig. 5b). Similarly to longitudinal modes, the cross-over from a free cantilever behavior to that of coupled cantilever does not increase with the mode number.

For constant periodic force acting on the cantilever, the boundary condition is

$$\Theta_0'(L) = \tilde{f}_{lat} \tag{34}$$

where $\tilde{f}_{lat} = F_{lat}H/c_T$ is the force acting on the cantilever. In this case, the cantilever response is

$$\Theta_0(L) = \frac{\tilde{f}_{lat} \tan \xi L}{\xi} \tag{35}$$

The resonant frequencies are determined by $\cos \xi L = 0$ or $\xi = (2n-1)\pi/L$.

## IV. Implications

In this section, we analyze the implications of frequency-dependent probe dynamics on PFM. In Section IV.1, electrostatic cantilever contribution to measured signal is analyzed. Frequency dependent dynamics of the cantilever is analyzed in Section IV.2. Section IV.3 discusses the relationship between the phase shift between domains and relative magnitude of electrostatic contribution, traditionally considered as a indication of quantitative imaging. Finally, Sections IV.4 and IV.5 discuss relative contributions of longitudinal and vertical signals to PFM contrast and approaches to absolute calibration of PFM response.

### IV.1. Cantilever contribution

The cantilever-surface capacitive force can be approximated as $F_{cap} = \varepsilon_0 S(V_{tip} - V_{surf})^2/2H^2$, where $S = Lw$ is the cantilever area.[53] The first harmonic of the distributed force is $q_0 = \varepsilon_0 S V_{ac} \Delta V/2LH^2$, where $H$ is the tip height (equal to the cantilever-surface separation) and $\Delta V = V_{dc} - V_{surf}$. The non-local contribution to the PFM signal can be conveniently rewritten in terms of the spring constant of the free oscillating cantilever, $k = Ewh^3/4L^3$ as $A_{nl} = -Lw\varepsilon_0 V_{ac} \Delta V / 48kH^2$. Thus, in the low frequency regime, the piezoresponse signal measured via local hysteresis loop measurements comprising of both electromechanical and non-local electrostatic parts is

$$PR = d_{eff} + \frac{Lw\varepsilon_0 V_{ac} \Delta V}{48kH^2}, \qquad (36)$$

where the first term is the sum of electromechanical and local electrostatic contributions and the second term is due to the cantilever buckling oscillations. Note that the non-local contribution is inversely proportional to the cantilever spring constant, while the electromechanical contribution is spring constant independent.

As discussed above, PFM imaging and quantitative piezoresponse spectroscopy requires the electromechanical interaction to be much stronger than that of the non-local electrostatic interaction. In the low frequency regime, this condition can be written as $k \gg Lw\varepsilon_0 \Delta V / 48 d_{eff} H^2$. Taking the estimates: $d_{eff}$ = 50 pm/V, $\Delta V$ = 5 V, $L$ = 225 μm, $w$ = 30 μm, $H$ = 15 μm, the condition on the spring constant is $k_{eff} >$ 0.55 N/m. This condition can be easily modified for cantilevers with different geometric properties and can be reformulated as a condition for the tip-surface potential difference for which the electromechanical contribution dominates. Note that while for zero tip-surface potential difference, $\Delta V = 0$, non-local interactions are formally absent, this condition is rarely achieved experimentally, unless a top-electrode experimental set-up is used.[54]

This electrostatic contribution also precludes the determination of electrostrictive constants from PFM hysteresis measurements because both electrostrictive and electrostatic contributions scale linearly with $\Delta V$ and cannot be distinguished unambiguously.

### IV.2. Frequency regimes in PFM

The relative contributions of electromechanical and electrostatic interactions to PFM contrast in the low frequency regime have been studied by Kalinin and Bonnell.[28] Here, we extend this analysis to the frequency dependence of the dominant contrast. Frequency dependence of the VPFM signal including electromechanical and local and non-local electrostatic contributions is given in Section II. The response is calculated for three cantilevers with different spring constants (Table 1). The tip height is taken to be $H = 15$ μm and $d_{eff} = 5$ pm/V. This corresponds to a weakly piezoelectric material with an electrostatic force of 14 nN/V acting on the tip. The spring constant of the tip-surface junction in all cases was taken as 1000 N/m (load ~100 nN) and assumed to be independent of the cantilever spring constant. This set of conditions corresponds to the case when the indentation force is dominated by capillary and adhesive forces, typical for imaging under ambient conditions.

Shown in Fig. 6 a,c,e are the amplitude maps for the PFM signal as a function of frequency and tip-surface potential difference, $\Delta V$, and calculated according to Eq. (28) for zero local electrostatic force, $f = 0$. A number of resonances (bright lines) and antiresonances (black lines) can be clearly seen. The phase changes by 180° across resonance and antiresonance lines. For low tip biases, the response is purely electromechanical and is independent of $\Delta V$. For higher DC biases, the response is dominated by non-local contributions and is linear in $\Delta V$. Note that the position of the resonances is determined solely by the cantilever properties and spring constant of the tip-surface junction and is independent of tip bias. Thus, the resonance frequency of the electrically driven cantilever in contact with the surface provides information only on the elastic properties of material, but not piezoelectric properties. At the same time, the zeroes on the response diagram are strongly bias dependent and therefore, the magnitude and frequency dependence of the nulling bias is related to the magnitude and sign of the electromechanical response.

The relative magnitudes of non-local and electromechanical contributions are illustrated in Fig. 6 b,d,f, illustrating the response map for $A_{piezo}/(A_{piezo}+A_{nl})$. The white region corresponds to dominant electromechanical contrast, while black regions correspond to dominant non-local electrostatic contributions. Note that in the low frequency limit the crossover between the two (indicated by an arrow) scales proportionally to the cantilever spring constant, as follows from the analysis in Section IV.1. At high frequencies, the relative contribution of the electromechanical contrast increases, indicative of dynamic cantilever stiffening. Also note that in the vicinity of the anti-resonances the non-local contribution is enhanced, while the resonances do not affect the relative contributions of these signals. Therefore, imaging at cantilever resonances will increase the signal to noise ratio, but will not affect the relative contributions of electrostatic and electromechanical responses, thus justifying the applicability of contact resonance-enhanced PFM imaging for low coercive bias materials.

In the presence of a non-zero local electrostatic force, the response diagram can be constructed as shown in Fig. 7. The intensity of red, green, and blue components represents the relative contributions of local electrostatic, non-local electrostatic, and electromechanical components. Note that the position of the boundary between the local electrostatic and electromechanical contribution is frequency independent and is determined only by the relative magnitudes of the piezo-coefficient, the electrostatic force, and the spring constant of the tip-surface junction. While these results are specific to the chosen parameters, similar diagrams can be readily constructed for different geometric parameters of the cantilever and the tip and material properties. However, in all cases, the qualitative features of Fig. 7 are valid, including the preponderance of the electromechanical response at low biases, the dominance of the non-local contribution in the vicinity of anti-resonances, the minimization of the non-local cantilever contribution at high frequencies, and that the frequency independent ratio between local electrostatic and electromechanical contributions is determined only by spring constant of tip-surface junction.

### IV.3. Phase shifts between antiparallel domain

One of the suggested guidelines for quantitative PFM imaging is the requirement for the phase shift between opposite domains to be 180°, with equal amplitudes on both sides of the domain wall. The simple formalism for analysis of the electrostatic contribution was established by Hong.[17] However, in the low frequency case, the phase shift between domains is always 180° or 0°, depending on the relative magnitudes of electrostatic and electromechnical contributions. In the dynamic case, the electromechanical and electrostatic contributions to the PFM signal have different phases and the response over $c^+$ and $c^-$ domains can be written as

$$PR_+ = d_{eff} + G_{elec}(V_{dc} - V_{av})\exp(i\psi) \qquad (37)$$

$$PR_- = -d_{eff} + G_{elec}(V_{dc} - V_{av})\exp(i\psi) \qquad (38)$$

where $\psi$ is the phase difference between electrostatic and electromechanical responses and $G_{elec}$ is electrostatic contribution to the signal including both local and non-local components. From Eqs. (37,38), the amplitude in the opposite domain is $PR_\pm = \sqrt{d_{eff}^2 + d_{elf}^2 \pm d_{eff} d_{el} \cos\varphi}$, where $d_{el} = G_{elec}(V_{dc} - V_{av})$ and the phase shift between the antiparallel domains is

$$\tan\varphi = \frac{2d_{el}(PR_- - PR_+)\sin\psi}{d_{el}^2(1 - \cos 2\psi) + 2PR_+ PR_-} \tag{39}$$

For a small electrostatic force contribution, $d_{el} \to 0$, the PFM amplitudes are $PR_\pm = d_{eff} \pm d_{el} \cos\varphi$. Whereas the phase difference between the domains is

$$\Delta\varphi = \pi - \sin(2\psi) d_{el}^2 / d_{eff}^2 . \tag{40}$$

Thus, the deviation of the phase shift between domains from 180° provides a measure of the electrostatic contribution, and validates its use as a criterion for quantitative PFM.

In the limit of a small piezoelectric contribution, $d_{eff} \to 0$, the domain contrast and phase change between the domains are respectively $PR_\pm = d_{el} \pm d_{eff} \cos\varphi$ and $\Delta\varphi = 2\sin(2\psi) d_{eff} / d_{ed} (\cos 2\psi - 3)$.

From Eqs. (37, 38) the following picture emerges. In the purely electromechanical case, $G_{elec}$ is identically zero. The response amplitudes are equal in $c^+$ and $c^-$ domain regions, while the phase changes by 180° between the domains. For domains with an arbitrary orientation, the absolute value of the amplitude signal provides a measure of the piezoelectric activity of the domain; in-plane domains or non-ferroelectric regions are seen as regions with zero response amplitude. PFM spectroscopy yields information on hysteresis behavior from which materials properties such as piezoelectric and electrostriction coefficients can be obtained. For a small non-local electrostatic contribution, $d_{eff} > G_{eloc}(V_{dc} - V_{av})$, the phase change between the domains will be less than 180° and the response amplitudes will no longer be equal in $c^+$ and $c^-$ domain regions. Non-ferroelectric regions will be seen as regions with small non-zero response amplitude. Finally, for a strong electrostatic contribution, the phase changes weakly between the domains, whereas the amplitude is a maximum for one orientation and a minimum for another.

### IV.4. Longitudinal and lateral contribution to PFM

As discussed in Section II, one of the problems in the interpretation of the vertical PFM signal is the contribution of longitudinal surface oscillations to the vertical PFM signal. This problem is closely related to the mechanism of signal transduction for the longitudinal and lateral components of surface displacement. In the low frequency regime with high indentation forces, the tip is expected to be coupled elastically to the surface. At the same time, for high frequencies and low indentation forces, the tip can slide along the surface, and tip motion is determined by the friction force between the two. The quantitative theoretical analysis of cantilever dynamics in this case is extremely complex and requires known phenomenological models for lateral contact mechanics of the tip-surface junction. In Section V, we present experimental results to illustrate this behavior.

### IV.5. PFM response calibration

One of the critical issues in PFM is the quantitative calibration of the vertical and lateral signal needed to obtain an electromechanical response vector. The signal collected from the lock-in amplifier is related to the surface displacement amplitude through a series of photodiode and lock-in sensitivities and gains, which are generally different for vertical and lateral PFM. One of the approaches for acquiring quantitative vertical PFM measurements includes calibration of photodiode sensitivity, which for example may include changing the deflection set-point in contact mode and measuring the associated change in surface position. Alternatively, the response can be quantified using a calibration sample (e.g. a quartz oscillator) with a known electromechanical response.[55] Calibration of lateral PFM presents a more complex problem. In principle, lateral sensitivity can be related to vertical sensitivity provided that the geometric parameters of the system are known. However, this approach is rather tedious. An alternative approach includes the use of a shear wave oscillator[33] in different orientations with respect to the cantilever axis to calibrate longitudinal and lateral contributions to PFM. However, the oscillators themselves are typically characterized by complex intrinsic dynamic behavior with a number of vertical and shear modes, which will be coupled with cantilever response.

## V. Experiment

A conventional way to access dynamic behavior in SPM is based on measurements of the amplitude and phase of the response as a function of frequency. In PFM, an additional contribution to the signal is given by the electrostatic force, which is linear in the tip-surface potential difference. To address PFM dynamics, we determine the vertical and lateral cantilever responses as a function of both tip dc bias and frequency, producing a 2D response diagram.

PFM was implemented on a commercial SPM system (Veeco MultiMode NS-IIIA) equipped with additional function generators and lock-in amplifiers (DS 345 and SRS 830, Stanford Research Instruments, and Model 7280, Signal Recovery). A custom-built sample holder was used to allow direct tip biasing and to avoid capacitive cross-talk in the SPM electronics. This holder does not contain a piezoactuator, thus reducing problems due to stray resonances. In most cases, the tip mount was glued to the holder using conductive epoxy to provide a rigid, electrically conductive connection.

A bias (both dc and ac) was applied directly to the tip either in contact mode (producing both an electrostatic and an electromechanical contribution) or at a small distance above the surface (yielding a purely electrostatic contribution). Measurements were performed on non-piezoelectric $SiO_2$, periodically poled $LiNbO_3$, and polycrystalline PZT ceramics using Pt and Au coated tips (NCSC-12 C, Micromasch, $l \approx 130$ μm, resonant frequency ~ 150 kHz, spring constant $k$ ~ 4.5 N/m), Co-Cr coated tips (Veeco, resonant frequency ~ 72 kHz, spring constant $k$ ~ 1 N/m) and calibrated Nanosensors tips.

## VI. Results and discussion

Experimental results on force-bias spectroscopy of cantilever dynamics are presented in Section VI.1. Bias effect on PFM imaging data is studied in Section VI.2. Tip wear effect of force-bias response spectrum is illustrated in Section IV.3 and approach for deconvolution of electrostatic and electromechanical components is suggested in Section VI.4. Finally, frequency dynamics of lateral PFM response is discussed in Section VI.5.

### VI.1. PFM dynamics in contact and non-contact modes

Fig. 8 illustrates the response diagrams acquired with an Co-Cr tip (tip 1) on a $SiO_2$ surface in contact and non-contact modes. The resonances in the contact regime are $\omega_1 =$ 407.95 ± 0.3 kHz, and $\omega_2 = 1075.020 \pm 0.15$ kHz. The ratio of the resonant frequencies is $\omega_2 : \omega_1 = 2.63 : 1$, as compared to theoretical ratio of 3.24:1. In the non-contact regime, resonances are $\omega_1 = 63.2 \pm 0.4$ kHz, $\omega_2 = 368.33 \pm 0.3$ kHz, and $\omega_3 = 1031.70 \pm 0.2$ kHz. From these data $\omega_3 : \omega_2 : \omega_1 = 16.32 : 5.83 : 1$, close to the theoretical ratio $17.55 : 6.23 : 1$. This is expected, since in the contact mode tip-surface interaction and absolute position of the tip along the cantilever significantly affects the resonance spectrum, while in the non-contact regime the behavior is close to the ideal. Note that the resonances, both in the contact and non-contact regimes, have zeroes at $V_{tip} = -81$ mV (contact) and $V_{tip} = 173$ mV (non-contact). The nulling bias is weakly frequency dependent, as expected, and the shift at high frequencies (> 1 MHz) can be attributed to the capacitive cross-talk in cabling. This behavior is more pronounced for weaker lateral response signals. Corresponding phase response diagrams illustrate that the phase changes by 180° across the zeroes in the bias direction and across the resonances in the frequency direction, resulting in a characteristic checkerboard pattern. For comparison, Fig. 8 c,d are the response diagrams from the LiNbO$_3$ surface. The resonant frequencies in the non-contact mode are identical to that for SiO$_2$, as illustrated in Fig. 9 a. Note that there are no zeroes in the bias interval of study and the phase is constant along the bias axis, suggesting that the surface is strongly charged ($V_{surf} > 4$ V). In the contact regime, the resonant frequencies are shifted compared to SiO$_2$ (by 8.2 kHz for $\omega_1$ and 4 kHz for $\omega_2$) due to the difference in elastic properties of the materials and the presence of an electromechanical response (compare to the inset in Fig. 9 b).

Shown in Fig. 10 are the vertical and lateral response diagrams for SiO$_2$ in non-contact (a,b) and contact (c,d) modes acquired using a Nanosensors tip (tip 2) ($h = 3$ μm, $w = 24$ μm, $L = 223$ μm, $k = 2.4$ N/m, as supplied by manufacturer). The general structure of the non-contact resonances is similar to that in Fig. 8, with the major resonances at $\omega_1 = 54.55 \pm 0.04$ kHz, $\omega_2$ = 346.73 ± 0.05 kHz, and $\omega_3$ = 980.55 ± 0.1 kHz. The ratio of the frequencies is $\omega_3 : \omega_2 : \omega_1 = 17.97 : 6.35 : 1$, very close to the theoretical ratio $17.55 : 6.23 : 1$. The frequency independent nulling bias is $V_{tip}$ = 1.0 V. The lateral response in Fig 10b shows strong resonances at $\omega_1$ = 54.6 ± 0.15, with a zero corresponding to $V_{tip}$ = 1.0 V. Close similarities are evident between resonances in the vertical and lateral modes suggesting that the latter is a result of the cross-talk between the normal and torsional cantilever oscillations. At higher frequencies, the lateral signal monotonically increases, presumably due to the cross-talk in the cabling. Response diagrams for the contact regime are shown in Fig. 10 c,d. The main resonances are similar to that for tip 1, however, an additional resonance at $\omega$ = 634.1 kHz emerges. The corresponding resonance frequency is sample independent. However, there is a nulling bias, suggesting that it is ultimately related to the electrostatic force.

Similarly to tip 1, the nulling bias is frequency independent in the contact regime. The lateral response shows features at the positions of the vertical resonances, indicative of cross-talk between normal and torsional modes. The frequency dependence of the vertical and lateral PFM signals in non-contact and contact regimes is summarized in Fig. 11a. A similar response diagram measured for PZT show a completely different behavior. The nulling bias is now strongly frequency dependent, as expected for the case when the relative contributions of electrostatic and electromechanical signals vary due to different frequency dependence (comp. Fig. 6 a,c,e). The deviation of nulling bias from surface potential value as a measure of electromechanical contribution to the signal, as discussed in detail below. The vertical and lateral PFM signals for PZT and $SiO_2$ are compared in Fig. 11 b. Note that the frequency dependence of the vertical response is drastically different, due to the difference in local elastic moduli and piezoelectric contributions to the signal. At the same time, the lateral PFM signal, expected to be zero for SiO$_2$ and non-zero for PZT, is virtually identical at high frequencies and differences are observed only at low frequencies.

**VI.2. Bias effect on PFM imaging**

To analyze the effect of bias conditions on the PFM signal, Fig. 12 illustrates the amplitude and phase images of polycrystalline PZT surface at biases of 0, -8, and 8 Volts. Note that, for the image close to nulling potential, the phase changes by 180° between domains. For a strong electrostatic contribution, the phase is purely positive or negative, as expected. The relative response amplitudes in grains 1 and 2 change with tip bias, in agreement with analysis in Section IV.3.

Further insight into the frequency-dependent dynamics of PFM can be obtained by comparison of the amplitude-frequency-bias response diagrams acquired from regions with dissimilar domain orientations. Fig. 13 shows such response diagrams for grain 1 (a,b) and grain 2 (c,d). Note that the orientation of the line corresponding to the frequency dependence of the nulling bias (vertical dark line) is opposite for these two grains, indicative of the opposite signs of the electromechanical contribution to the PFM signal for antiparallel grains.

This analysis allows the frequency range for optimal PFM imaging to be established. Indeed, for purely electrostatic imaging, the nulling bias is equal to the surface potential. In the electromechanical limit, the response is bias independent, and there is no nulling bias. In the intermediate case, the nulling bias is $V_{null} = V_{surf} \pm d_{eff}/G_{elec}$ (the sign corresponds to the domain orientation) and depends on the relative magnitudes of electrostatic and electromechnical contributions. Therefore, the frequencies for which $|V_{null}|$ is maximal correspond to the frequencies at which the electromechanical contribution is dominant, and the resulting PFM image has optimal contrast.

### VI.3. Tip wear effect on response diagram

This analysis suggests that the frequency dependence of the nulling bias, $V_{null}(\omega)$, provides information on the relative magnitude of the electromechnical contribution to the PFM signal. As an illustration, Fig. 13 c,e,g illustrates the response diagrams for grain 2 acquired after several repetitive topographic scans. Note that $V_{null}$ is strongly bias dependent for a good tip. For a deteriorated tip, the electromechanical contribution to the signal is small, so that $V_{null} \approx V_{surf}$, and is now virtually bias independent. Therefore, the degree of deviation of the nulling bias from a constant, for a sample with known and appreciable piezoelectric properties, provides a measure of tip quality.

### VI.4. Deconvolution of the electrostatic and electromechanical contributions

The amplitude and phase response diagrams allow the electromechanical and electrostatic contributions to the PFM signal to be distinguished. Briefly, the PFM $x$-signal, defined as $PR = A\cos\varphi$, can be represented as

$$PR_+ = \tilde{d}_{eff} + \tilde{G}_{elec}(V_{dc} - V_{surf}), \tag{41}$$

where $\tilde{d}_{eff}$ and $\tilde{G}_{elec}$ are the electromechanical and electrostatic contributions now including a frequency-dependent phase multiplier. From Eq. (41), the electrostatic contribution to the PFM signal can be determined from the slope, $\tilde{G}_{elec} = c$, of the response vs. bias curve at each frequency. The electromechanical contribution is related to the intercept, $b$, as $\tilde{d}_{eff} = b + cV_{surf}$. Note that while the electrostatic contribution can be determined unambiguously, the electromechanical contribution depends on a known surface potential, $V_{surf}$, which can be determined e.g. from non-contact measurements.

Shown in Fig. 14 are the bias and frequency dependent response diagrams for non-piezoelectric $SiO_2$ in non-contact (a) and contact (c) regimes, as compared to PZT grain 2 acquired with good (e) and deteriorated (g) tips. The function $y = b + cV_{tip}$ was fit to the signal for each frequency. Shown in Fig. 14 b,d,f,h are error maps defined as $PR - (b + cV_{tip})$, representing the deviation of the actual response from a purely linear response. The scale for Fig. 14 b,d,f,h is 1% of full scale for Fig. 14 a,c,e,g. Note that the deviation from linearity is extremely small, suggesting the validity of Eq. (41). The maximum deviations are observed in the vicinity of resonances, where non-linear effects become pronounced.

The frequency dependences of the electromechanical and electrostatic responses for these materials are shown in Fig. 15 a,b. Note that the electromechanical response is greatest for PZT grain 1 with a good tip, slightly smaller for grain 2, and is negligibly small for $SiO_2$ in the non-contact and contact modes (Fig. 15 a), as expected. In comparison, the electrostatic response is comparable for all materials (Fig. 15 b). The resonant frequencies for the electromechanical and electrostatic signals coincide for a given sample, as predicted by Eq. (5). An alternative approach for distinguishing electrostatic and electromechanical contributions has been suggested by Harnagea[21] based on measurements of an amplitude-frequency curve of piezoelectric and non-piezoelectric materials. However, this approach is applicable only if the resonant frequencies of a cantilever in contact with the surface are identical, which is not the case for dissimilar materials. The differentiation of these contributions based on the bias dependence of the response provides a more rigorous approach provided that the surface potential is known.

Analysis of the relative electromechanical vs. electrostatic contribution far from the resonances at high frequencies presents a more challenging problem. Intrinsic contributions (e.g. due to the electrocapillary effect on the tip surface junction or extrinsic contributions such as capacitive cross-talk in cabling and the detector), can result in a shift of the nulling bias from the surface potential (e.g. the weak frequency dependence of the nulling bias in Fig. 10e for $SiO_2$), resulting in a non-zero effective piezoresponse even for non-piezoelectric materials. Fig. 15 c reveals the frequency dependence of the electromechanical contribution for PZT and $SiO_2$. While for most frequencies the effective electromechanical response for PZT is ~2 orders of magnitude higher than that for $SiO_2$, the later is non-zero, thus necessitating further studies. Also shown in Fig. 15 d is the frequency dependence of the ratio of the electromechanical and electrostatic responses for PZT and $SiO_2$. The position of the minima at anti-resonances is in agreement with the interpretation of Fig. 6,7.

### VI.5. Lateral PFM

Finally, experimental data in Fig. 11 illustrates that the frequency behavior of the lateral signal differs strongly from that of the vertical. While in the latter case a number of cantilever resonances are observed in the frequency range of study (0-2 MHz), the frequency dependence of the lateral signal is relatively featureless (Fig. 11 a,b). At high frequencies, the response is dominated primarily by capacitive cross-talk in the detector electronics and cabling. The differences in the response of PZT and $SiO_2$ and in the lateral response diagrams of different PZT grains is observed only at frequencies << 100 kHz. The lateral PFM image contrast was found to decay quickly at frequencies above ~10 kHz (not shown), in agreement with results by Kholkin.[56] These results suggest that, for lateral PFM, effective detection is possible only at low (<10 kHz) frequencies. This response decays rapidly as frequency increases, presumably due to the onset of sliding friction between the tip and the surface. These results also suggest that the longitudinal contribution to PFM contrast will be absent at high frequencies, due to similarities in vibration transduction mechanisms for longitudinal and lateral components of in-plane surface displacement. These conclusions are corroborated by measurements performed using vertical and shear oscillators, to be reported elsewhere.[57] Therefore, quantitative measurements of vertical and lateral PFM signals require imaging at different frequencies – the low frequency regime favors lateral PFM measurements, while imaging at high frequency provides vertical PFM data free from a longitudinal contribution.

## VI. Conclusions

The dynamic cantilever behavior in vertical and lateral PFM is analyzed using a combination of theoretical modeling and experimental response-frequency-bias spectroscopy. In general, the vertical PFM signal has contributions from the normal component of the electromechanical response vector, the longitudinal component of the response vector, the local electrostatic force acting on the tip, and the distributed electrostatic force acting on the cantilever. Lateral PFM is dominated by the lateral surface displacement resulting in torsional cantilever oscillations.

In vertical PFM, the positions of the contact resonances are determined solely by the cantilever spring constant and the effective spring constant of the tip-surface junction. Thus, the resonance frequency for the electrical excitation allows mapping of the local elastic properties of the surface, similar to atomic force acoustic microscopy with mechanical excitation. The relative magnitudes of the non-local and local electrostatic contributions and normal surface displacement are additive, and frequency-bias response diagrams have been constructed to select the region with dominant responses. It is shown that the non-local contribution decays faster with frequency than does the electromechanical and local electrostatic contributions. The relative non-local contribution is maximum at the anti-resonances, but does not change significantly at the resonances, validating the use of resonance enhancement for PFM imaging. The relative contribution of electromechanical and local electrostatic contributions is frequency independent and is determined by the tip bias, tip surface capacitance, and the effective spring constant of tip-surface junction only. Quantitative electromechanical measurements require imaging at low tip biases or (for PFM spectroscopy) the use of high indentation forces and high spring constant cantilevers. The resonance frequencies for vertical and longitudinal vibration modes are generally different, suggesting that all three components of the electromechanical response vector can be mapped by suitable choices of imaging frequencies.

The longitudinal and lateral contributions to PFM are governed by the lateral contact mechanics of the tip-surface junction. For high frequencies or low indentation forces, the onset of sliding friction between the tip and the surface will minimize the transmission of in-plane surface vibrations to the tip. This suggests that lateral PFM imaging is optimal at relatively low frequencies, when the vibration transfer from surface to the tip is effective. Conversely, vertical PFM is optimal at relatively high frequencies, when the longitudinal contributions are minimal.

Frequency-bias response diagrams are shown to be a convenient tool for analysis of image formation mechanisms in PFM. In particular, the frequency dependence of the nulling bias provides information on the electromechanical contribution to the signal and delineates frequency regimes for optimal quantitative electromechanical imaging. The electrostatic contribution to the PFM signal can be determined unambiguously as the slope of the response-bias curve at each frequency. The electromechanical contribution can be determined from the intercept; however, knowledge of the surface potential is required for unambiguous evaluations. The response curve measurements on piezoelectric and non-piezoelectric materials do not provide this information due to the shifts in relative positions of resonance peaks. Finally, observations of bias-dependent resonances in lateral response measurements indicate coupling to vertical modes, and lateral PFM measurements must be performed at frequencies far from the vertical resonances of the cantilever.

Support from ORNL SEED funding under Contract DE-AC05-00OR22725 is acknowledged (SVK).

Table 1

Cantilever Properties

| Cantilever | $w$, μm | $h$, μm | $L$, μm | $k$, N/m |
|---|---|---|---|---|
| A | 24 | 8 | 224 | 45 |
| B | 24 | 3 | 224 | 2.4 |
| C | 24 | 1.05 | 224 | 0.1 |

# Figure captions

**Fig. 1.** Schematic diagram of non-local cantilever effects in PFM. Displacement of a laser beam induced by cantilever deflection (a) is equivalent to that due to cantilever buckling induced by a uniformly distributed load (b). (c) Electrostatic contribution to PFM contrast. A local electrostatic force between the tip and the surface results in a vertical tip displacement. (d) Electromechanical contrast in PFM due to field-induced strain in the material (inverse piezoelectric effect).

**Fig. 2.** (a) The electromechanical response of the surface in the general case is a vector having components normal, longitudinal in-plane (along the cantilever axis) and lateral in-plane (perpendicular to cantilever axis). (b) The longitudinal response contributes to the cantilever bending angle, resulting in a measured normal displacement.

**Fig. 3.** Equivalent circuit for the flexural cantilever oscillations in vertical PFM.

**Fig. 4.** (a) Simplified equivalent circuit for the normal oscillations of the cantilever coupled elastically to the surface and (b) resonant frequency as a function of normalized tip-surface spring constant. Simplified equivalent circuit for the longitudinal oscillations of the cantilever for the case of (c) linear elastic coupling to the surface and (d) frictional force coupling. (e) Resonant frequency as a function of the normalized tip-surface spring constant for case (c).

**Fig. 5.** Simplified equivalent circuit for the torsional oscillations of the cantilever for the case of (a) linear elastic coupling to the surface and (b) friction force coupling. (c) Resonant frequency as a function of the normalized tip-surface spring constant for case (a).

**Fig. 6.** Frequency and bias dependent amplitude response diagram (a,c,e) and the regions of dominant electromechanical contribution (b,d,f) of the cantilevers A (a,b), B (c,d) and C (e,f). Cantilever parameters are summarized in Table I. Plotted is log(Amplitude) (a,c,e). White corresponds to the regions with a dominant electromechanical contribution, while black corresponds to regions with a dominant non-local electrostatic contribution (b,d,f).

**Fig. 7.** Frequency and bias dependent amplitude response diagrams when electromechanical, local, and non-local electrostatic contributions are present. Blue corresponds to dominant electromechnical response, red to local electrostatic, and green to non-local electrostatic responses. Shown are diagrams for cantilever A for $f = 14$ N/V (a), cantilever A for $f = 0.7$ N/V (b), cantilever B for $f = 14$ N/V (c) and cantilever C for $f = 14$ N/V (d).

**Fig. 8.** Frequency-bias amplitude (a,c,e,g) and phase (b,d,f,h) response diagrams for $SiO_2$ (a-d) and $LiNbO_3$ (e-h) in contact (a,b,e,f) and non-contact (c,d,g,h) regimes. Vertical scaling presented as log (Amplitude) (a,c,e,g).

**Fig. 9.** Frequency dependence response amplitude for in non-contact (a) and contact (b) regimes for $LiNbO_3$ (solid line) and $SiO_2$ (dash). Insets show an expanded view at the position

of the first resonance peak. Note that resonances coincide in the non-contact and differ in the contact regimes due to a difference in the elastic constant of tip-surface junction.

**Fig. 10.** Frequency-bias amplitude (a,c,e,g,i,k) and phase (b,d,f,h,j,l) response diagrams for $SiO_2$ in the non-contact mode (a-d), $SiO_2$ in the contact mode (e-h) and PZT in the contact mode (j-l). Shown are the vertical (a,b,e,f,i,j) and lateral (c,d,g,h,k,l) response diagrams.

**Fig. 11.** (a) Frequency dependence of the vertical and lateral response amplitudes for $SiO_2$. Shown are vertical response in contact mode (solid), lateral response in contact mode (dash), vertical response in non-contact mode (dash-dot), and lateral response in non-contact mode (dash-dot-dot). (b) Frequency dependence of the vertical and lateral response amplitudes for $SiO_2$ and PZT in contact mode. Shown are the vertical response for $SiO_2$ (solid), PZT (dash-dot), lateral response for $SiO_2$ (dash), and PZT (dash-dot-dot).

**Fig. 12.** Amplitude (a,c,e) and phase (b,d,f) piezoresponse images of a PZT surface at a tip bias of 0 V (a,b), 8V (c,d) and -8 V (e,f). Note that the phase changes by 180° between anti-parallel domains in (b), however, for predominantly electrostatic contributions, the phase is constant (f). The relative response amplitude inverts with tip bias (compare domains 1,2,3 in c,e).

**Fig. 13.** Amplitude response map in grain 1 (a) and grain 2 (b) in Fig. 12. Note the difference in the position of the anti-resonance curve, indicative of the opposite sign of the electromechanical contribution to the signal for anti-parallel grains. Arrows correspond to frequencies at which the electromechanical contribution dominates. Amplitude response map on grain 2 captured at different stages of tip degradation: after 1 (c) and 3 (d) successive scans.

**Fig. 14.** Piezoresponse x-signal (a,c,e,g) for $SiO_2$ in the non-contact regime (a), contact regime (c), PZT with a good tip (e) and a deteriorated tip (g). Also shown are corresponding error maps after the subtraction of the linear contribution (b,d,f,h). Note, the scale of the error maps is 1% of the scale for response maps.

**Fig. 15.** Frequency dependence of (a) electromechanical and (b) electrostatic response for PZT with a good tip (solid), PZT with a deteriorated tip (dash), $SiO_2$ in the non-contact mode (dash dot), and contact mode (dash dot dot). (c) Frequency dependence of the piezoresponse signal for PZT (solid) and $SiO_2$ (dash) on a logarithmic scale. (d) Ratio of the electromechanical and electrostatic responses for PZT (solid) and $SiO_2$ (dash).

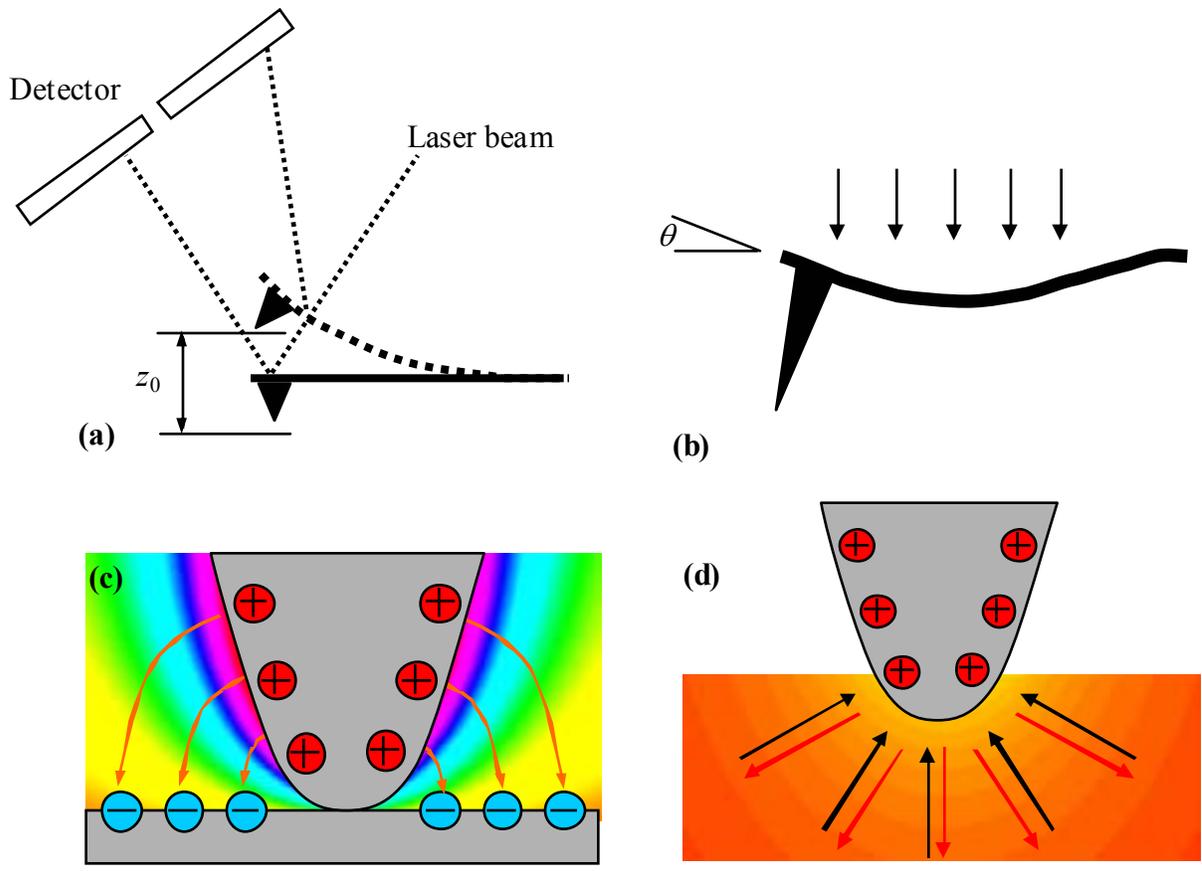

**Figure 1.** S. Jesse, A.P. Baddorf, and S.V. Kalinin

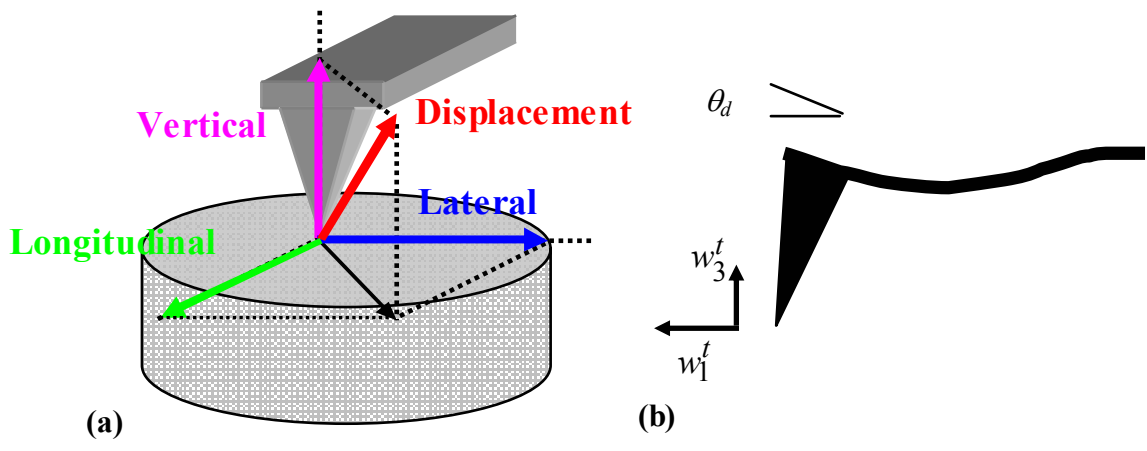

**Figure 2.** S. Jesse, A.P. Baddorf, and S.V. Kalinin

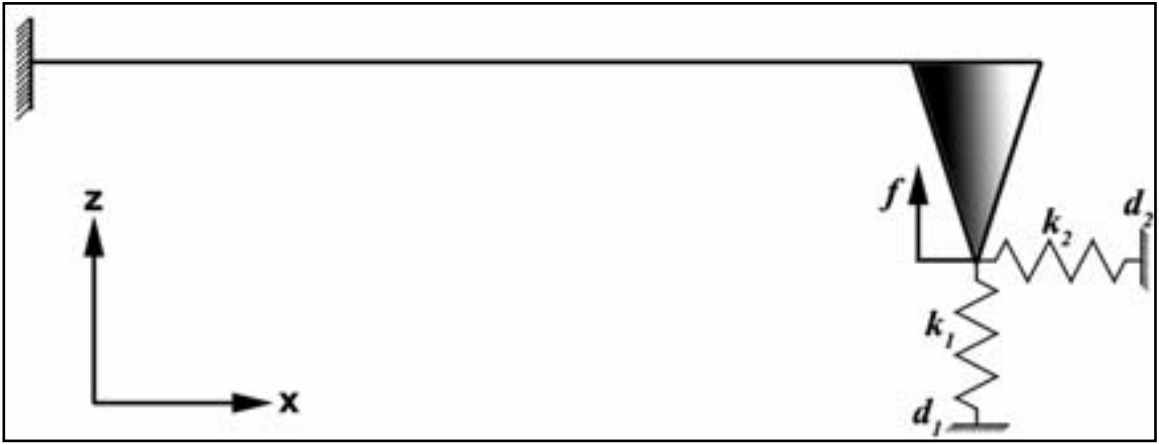

**Figure 3.** S. Jesse, A.P. Baddorf, and S.V. Kalinin

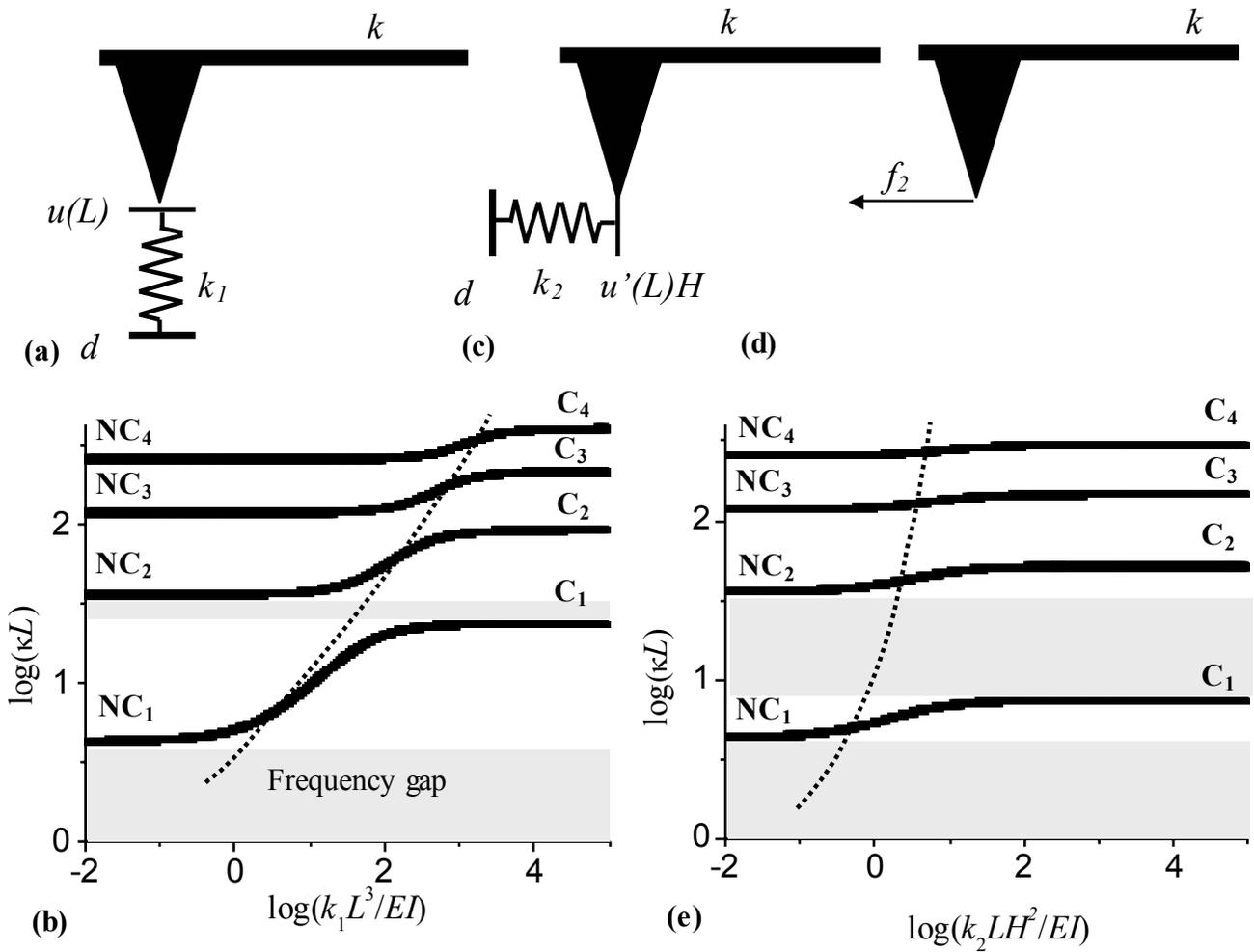

**Figure 4.** S. Jesse, A.P. Baddorf, and S.V. Kalinin

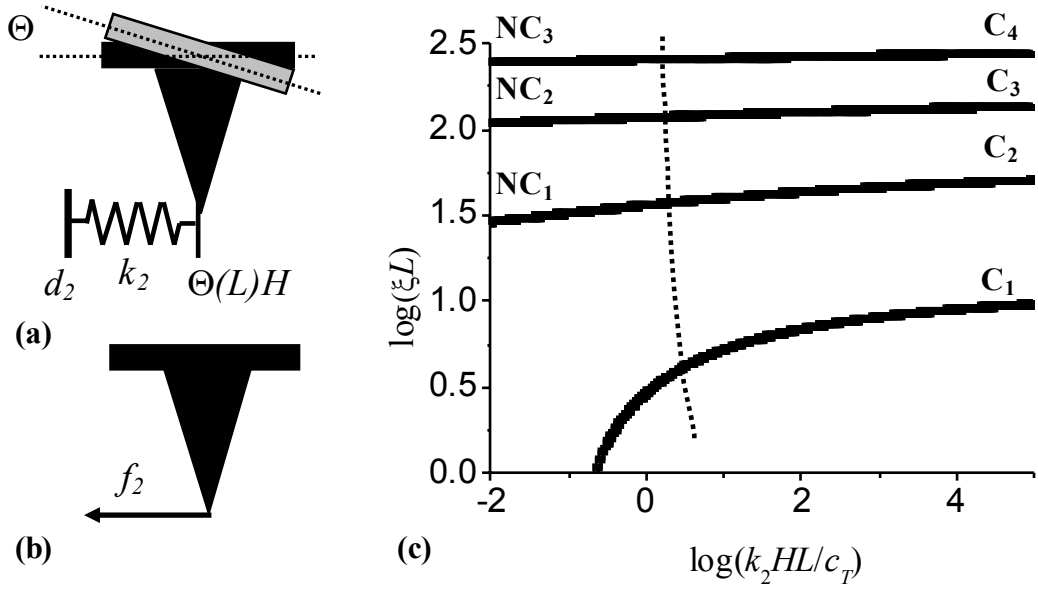

**Figure 5.** S. Jesse, A.P. Baddorf, and S.V. Kalinin

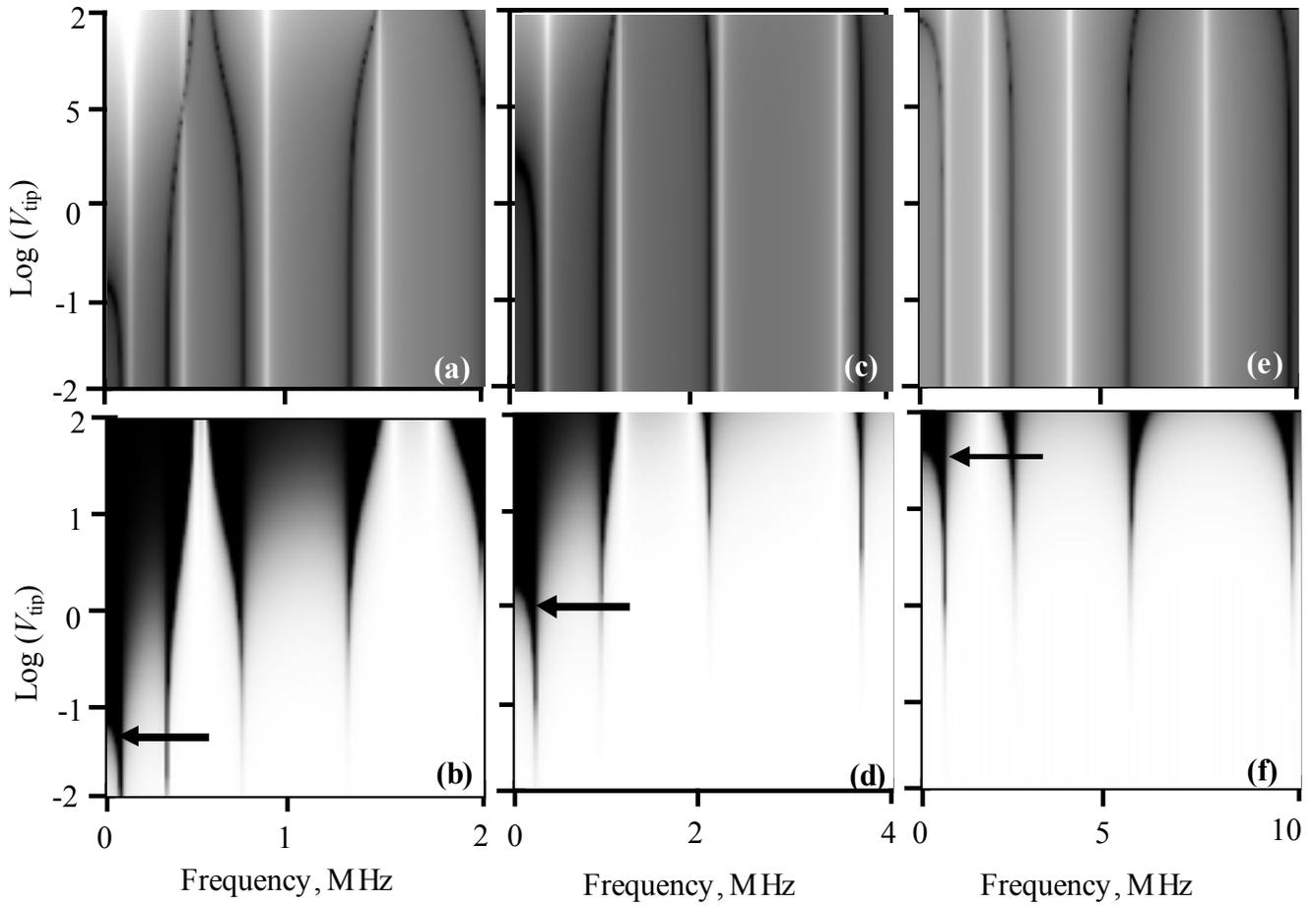

**Figure 6.** S. Jesse, A.P. Baddorf, and S.V. Kalinin

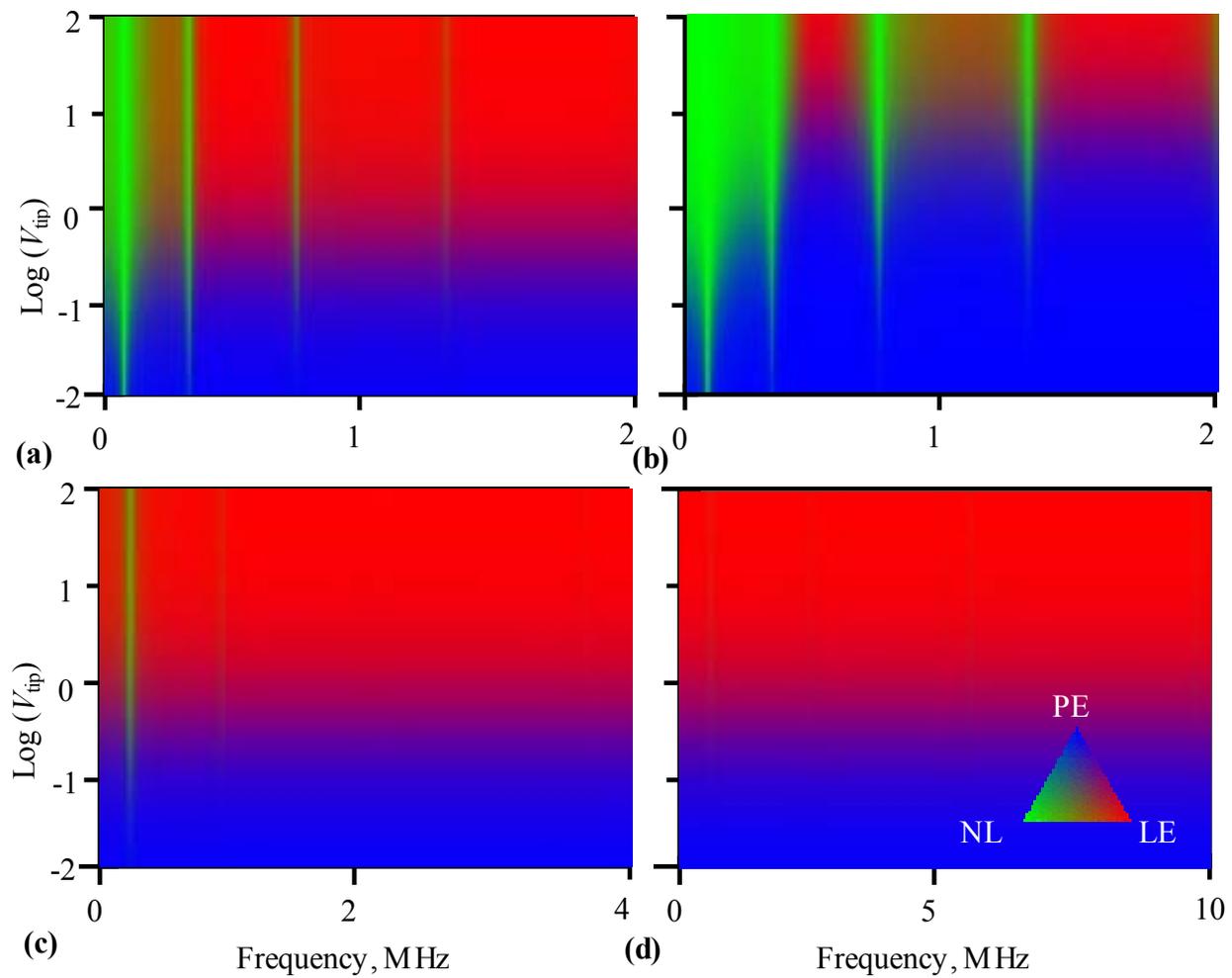

**Figure 7.** S. Jesse, A.P. Baddorf, and S.V. Kalinin

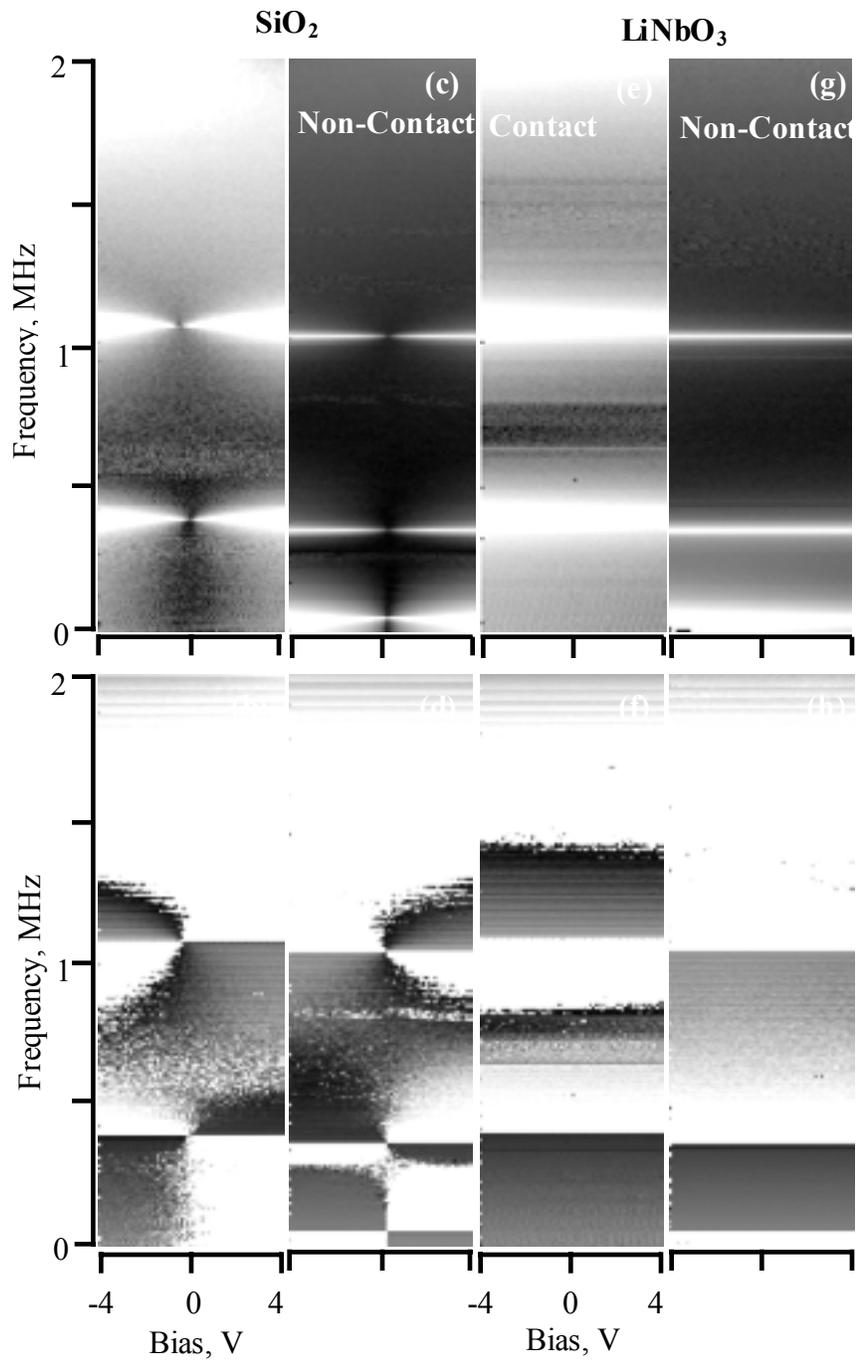

**Figure 8.** S. Jesse, A.P. Baddorf, and S.V. Kalinin

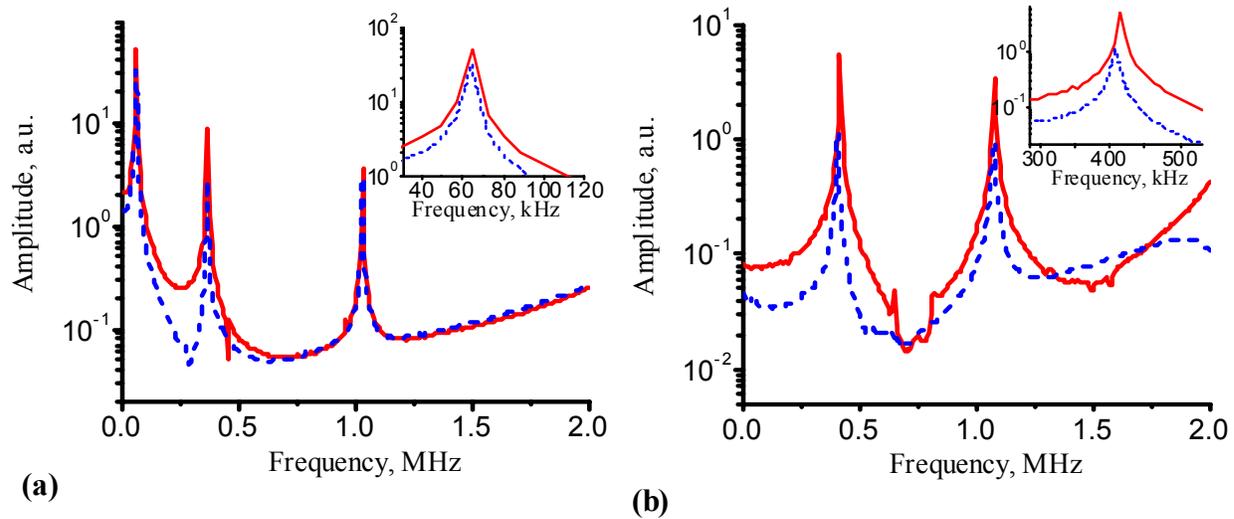

**Figure 9.** S. Jesse, A.P. Baddorf, and S.V. Kalinin

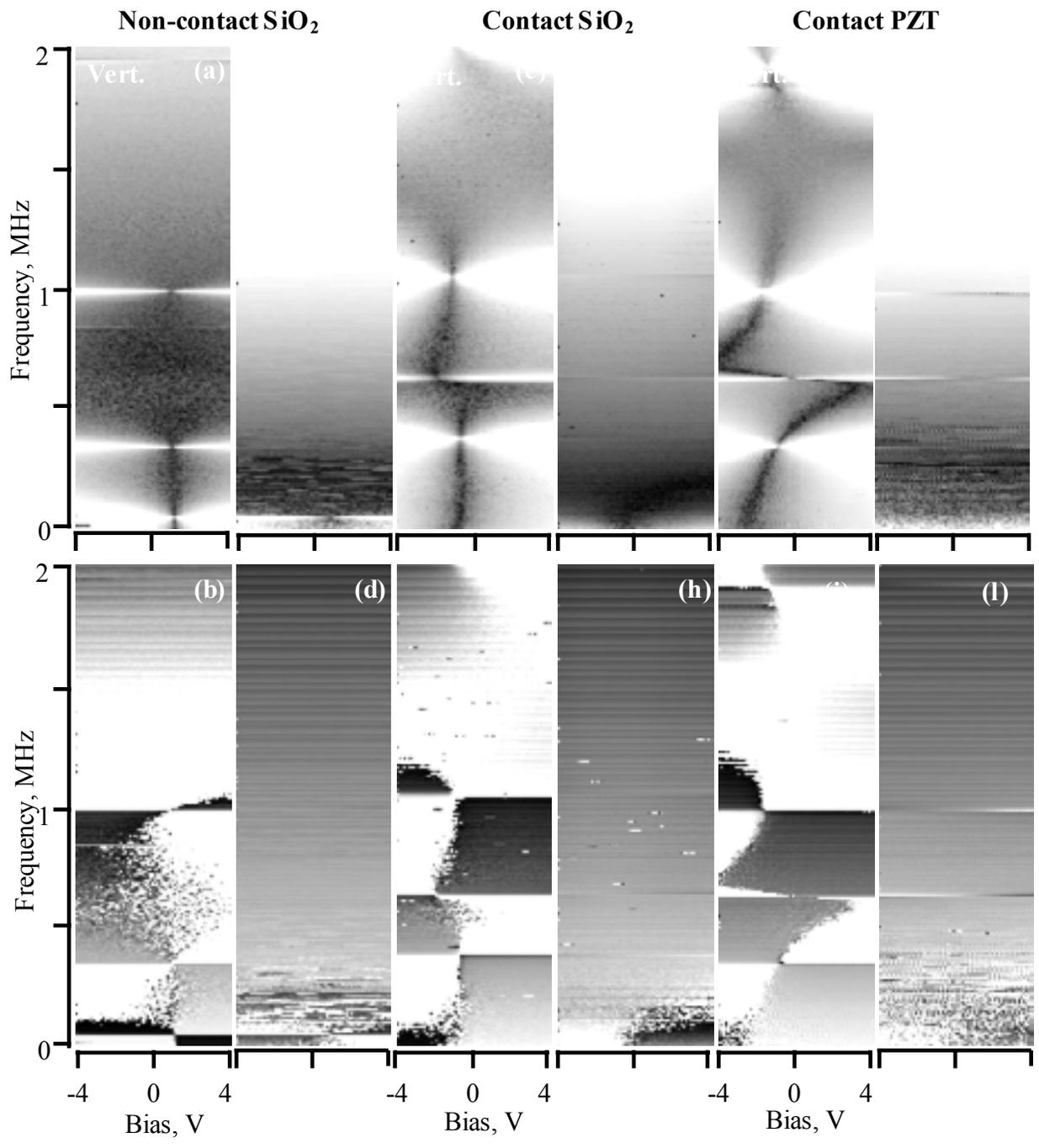

**Figure 10.** S. Jesse, A.P. Baddorf, and S.V. Kalinin

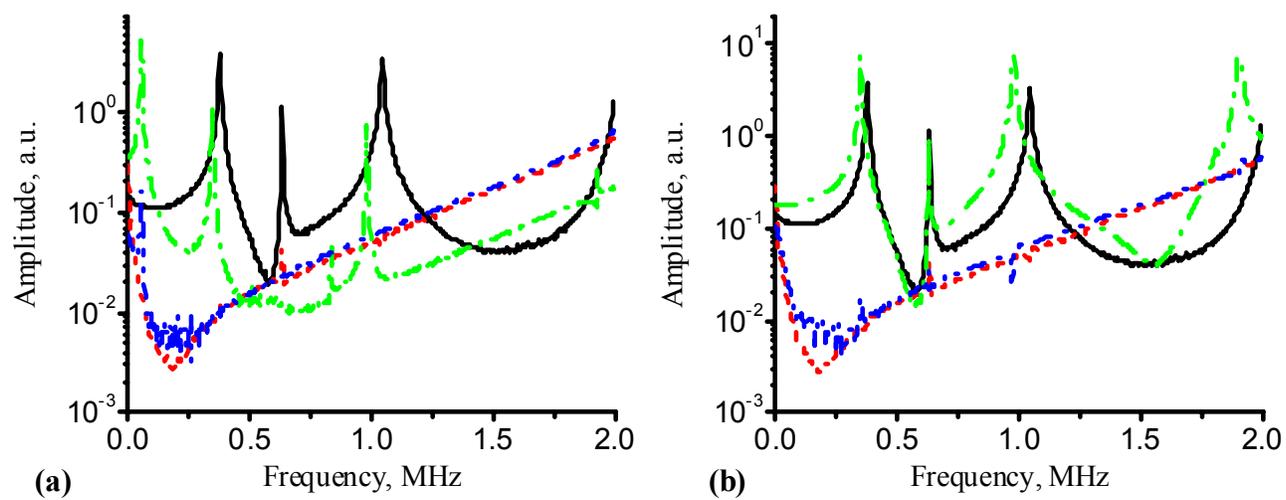

**Figure 11.** S. Jesse, A.P. Baddorf, and S.V. Kalinin

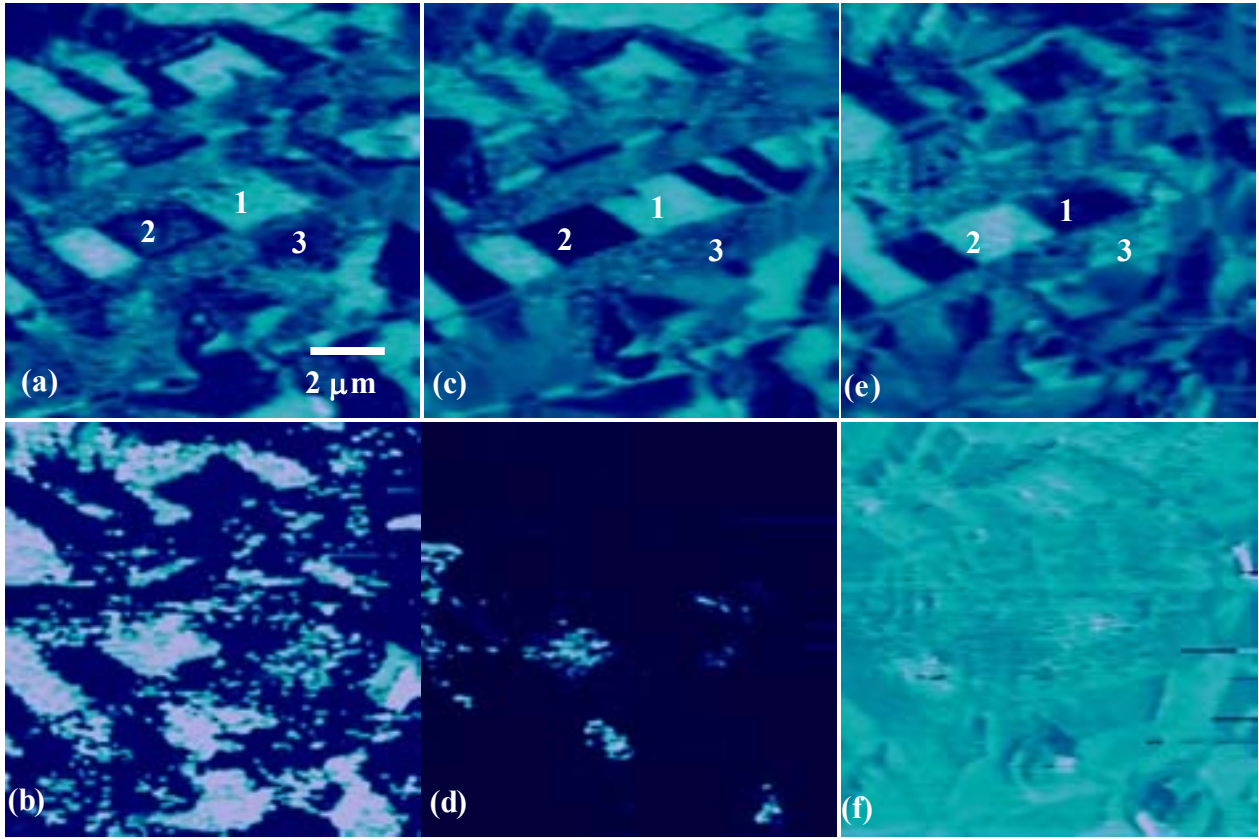

**Figure 12.** S. Jesse, A.P. Baddorf, and S.V. Kalinin

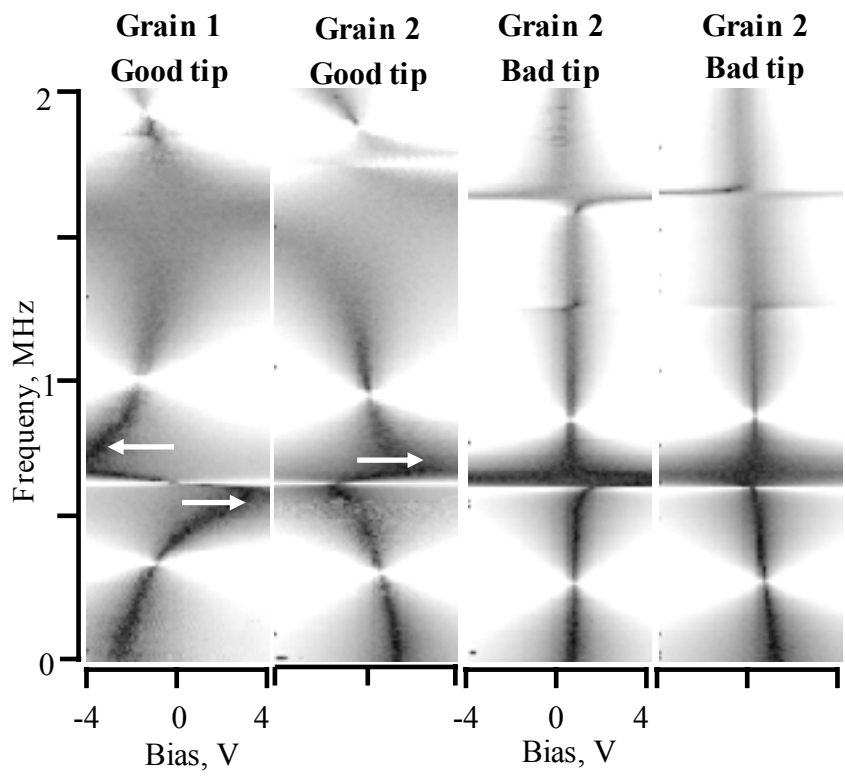

**Figure 13.** S. Jesse, A.P. Baddorf, and S.V. Kalinin

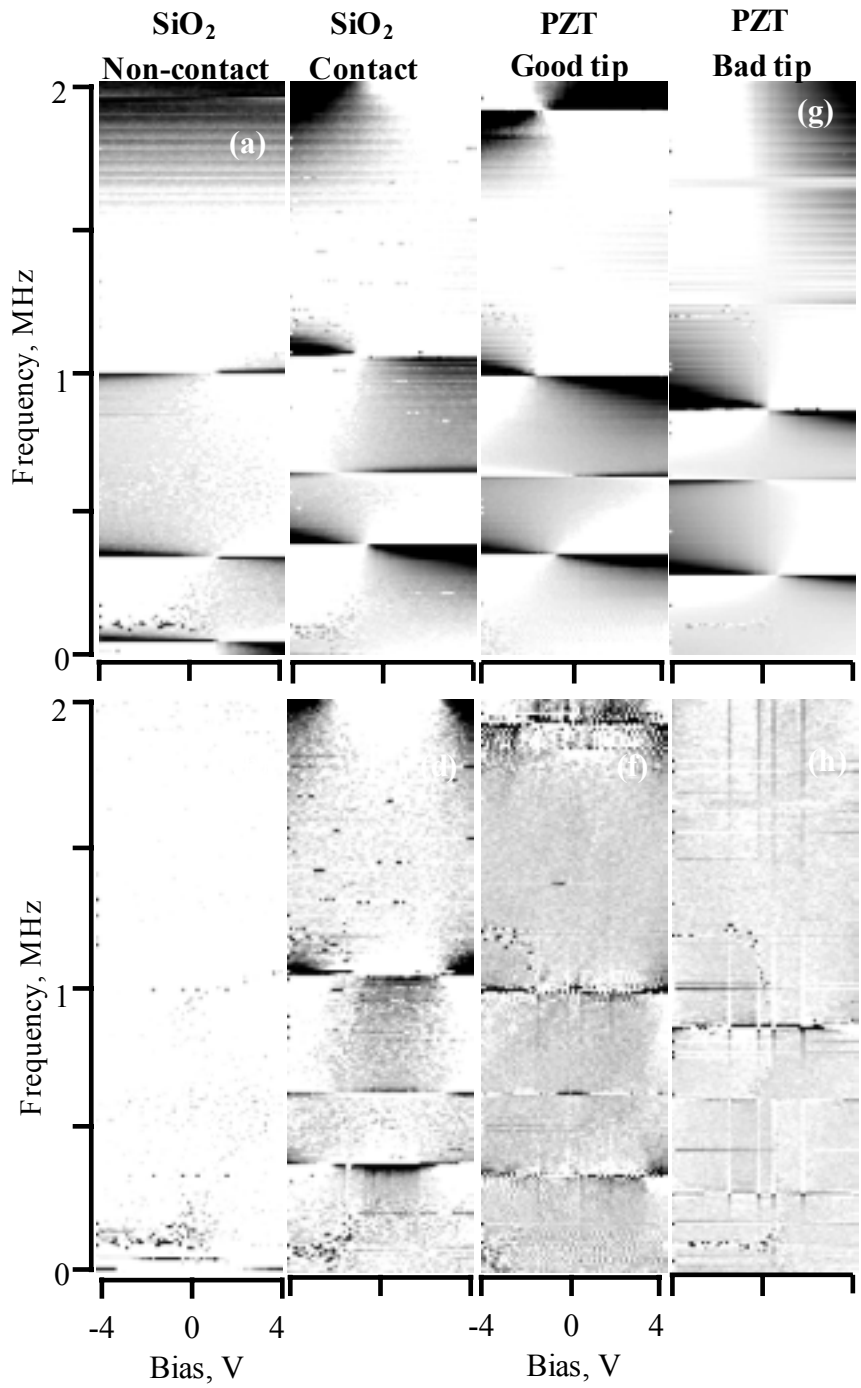

**Figure 14.** S. Jesse, A.P. Baddorf, and S.V. Kalinin

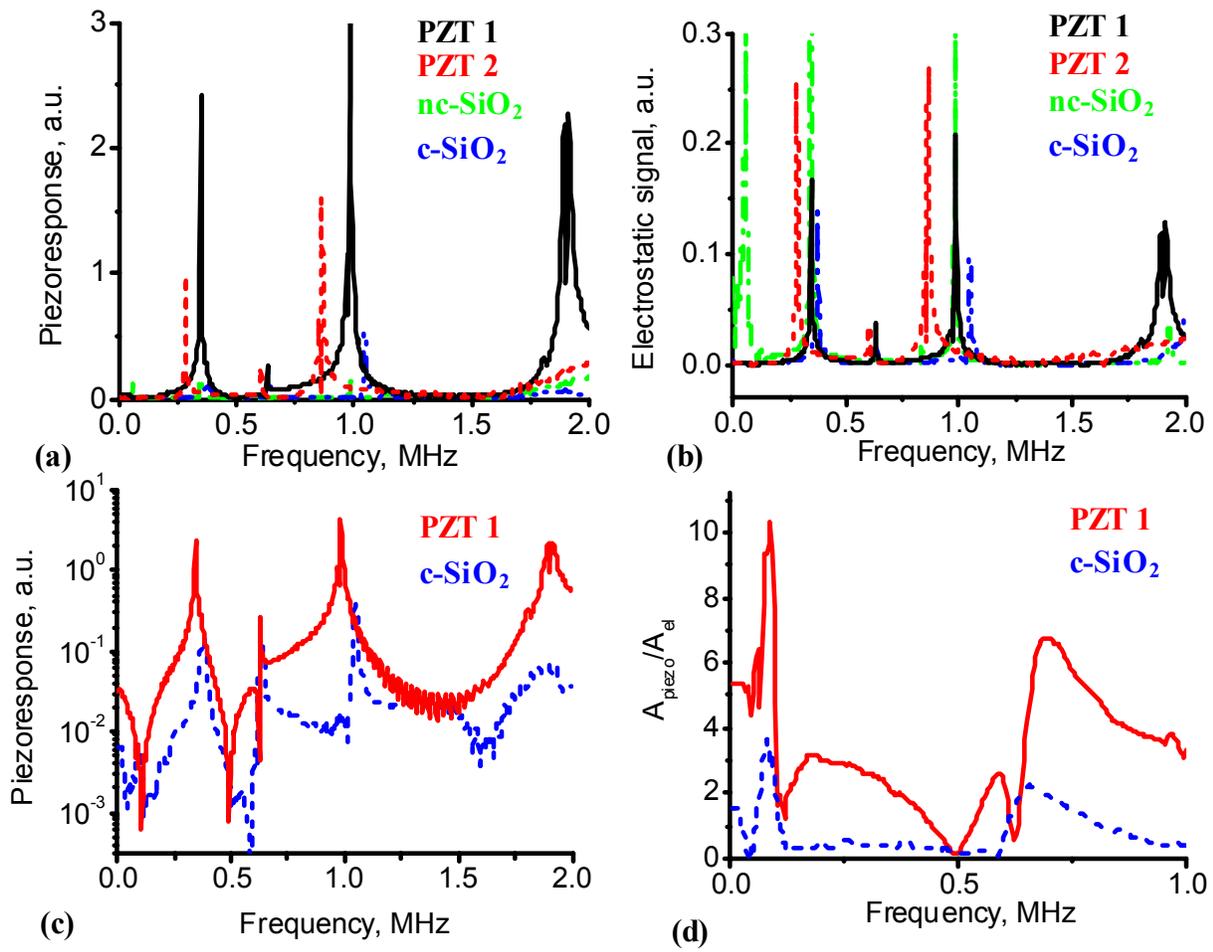

**Figure 15.** S. Jesse, A.P. Baddorf, and S.V. Kalinin

[54] A. Gruverman, in *Encyclopedia of Nanoscience and Nanotechnology*, edited by H.S. Nalwa, Vol. 3, pp.359-375 (American Scientific Publishers, Los Angeles, 2004).

[55] C. Ganpule, *Nanoscale Phenomena in Ferroelectric Thin Films*, Ph.D. thesis, University of Maryland, College Park (2001).

[56] I.K. Bdikin, V.V. Shvartsman, S.-H. Kim, J.M. Herrero, and A.L. Kholkin, Mat. Res. Soc. Symp. Proc. **784**, C11.3 (2004).

[57] S. Jesse et al, to be published

65